\documentclass[12pt,aps,tightenlines,notitlepage,showpacs,superscriptaddress]{revtex4-1}
\usepackage[dvips]{epsfig}

\usepackage{amsmath}
\usepackage{amssymb}
\usepackage{bm}
\usepackage[mathscr]{eucal}
\usepackage{theorem}
\usepackage{graphicx}
\usepackage{psfrag}
\usepackage{subfigure}
\usepackage{color}
\usepackage{wasysym}
\include{graphix}
\usepackage{framed}
\usepackage{enumitem}%
\usepackage{lipsum}
\usepackage{float}
\usepackage{MnSymbol}

\DeclareMathAlphabet{\mathpzc}{OT1}{pzc}{m}{it}

\graphicspath{{figs/}{figs_pdf/}}

\newcommand{\imag}{\mathrm{i}}
\newcommand{\mathe}{\mathrm{e}}
\newcommand{\mathd}{\mathrm{d}}

\newcommand{\myvar}{\sigma}

\newcommand{\mylep}{\Lambda}
\newcommand{\vecx}{\bm{x}}
\newcommand{\vecr}{\bm{r}}
\newcommand{\vecv}{\bm{v}}

\newcommand{\veck}{\bm{k}}

\newcommand{\mydiff}{\mathcal{D}}
\newcommand{\gammanondim}{\mathcal{C}_{\mathrm{n}}}
\newcommand{\myq}{\mathpzc{q}}

\begin{document}

\bibliographystyle{apsrev}

\title{Flow-parametric regulation of shear-driven phase separation\\
in two and three dimensions}

\author{Lennon \'O N\'araigh}
\email{onaraigh@maths.ucd.ie}

\author{Selma Shun}
\affiliation{School of Mathematical Sciences, University College Dublin, Belfield, Dublin 4}
\affiliation{Complex and Adaptive Systems Laboratory, University College Dublin, Belfield, Dublin 4}

\author{and Aurore Naso}
\affiliation{Laboratoire de M\'ecanique des Fluides et d'Acoustique, CNRS, \'Ecole Centrale de Lyon, Universit\'e de Lyon, 69134 \'Ecully Cedex, France}

\date{\today}

\begin{abstract}
The Cahn--Hilliard equation with an externally-prescribed chaotic shear flow is studied in two and three dimensions.  The main goal is to compare and contrast the phase separation in two and three dimensions, using  high-resolution numerical simulation as the basis for the study.    The model flow is parametrized by its amplitudes (thereby admitting the possibility of anisotropy), lengthscales, and multiple time scales, and the outcome of the phase separation is investigated as a function of these parameters as well as the dimensionality.
In this way, a parameter regime is identified wherein the phase separation and the associated coarsening phenomenon are not only arrested but in fact the concentration variance decays, thereby opening up the possibility of describing the dynamics of the concentration field using the theories of advection diffusion.  This parameter regime corresponds to long flow correlation times, large flow amplitudes and small diffusivities.  The onset of this hyperdiffusive regime is interpreted by introducing Batchelor lengthscales. A key result is that in the hyperdiffusive regime, the distribution of concentration (in particular, the frequency of extreme values of concentration) depends strongly on the dimensionality. 
Anisotropic scenarios are also investigated: for scenarios wherein the variance saturates (corresponding to coarsening arrest), the direction in which the domains align depends on the flow correlation time.  Thus, for correlation times comparable to the inverse of the mean shear rate, the domains align in the direction of maximum flow amplitude, while for short correlation times, the domains initially align in the opposite direction.  However, at very late times (after the passage of thousands of correlation times), the fate of the domains is the same regardless of correlation time, namely alignment in the direction of maximum flow amplitude.  A theoretical model to explain these features is proposed.   These features and the theoretical model carry over to the 3D case, albeit that an extra degree of freedom pertains, such that the dynamics of the domain alignment in 3D warrant a more detailed consideration, also presented herein.
%

\end{abstract}


\maketitle

\section{Introduction}

When an alloy or a binary mixture in which both components are initially uniformly present undergoes rapid cooling below a critical temperature, both phases separate. In the absence of flow, domains of higher concentration in one or the other phase form and grow algebraically in time, obeying a coarsening process referred to as spinodal decomposition.
In fluids, this phase separation is often more complex than in solid alloys because of the interaction between hydrodynamic and coarsening effects~\cite{Siggia79}. Even when the flow is externally driven, the situation is  complicated. In the presence of shear flow~\cite{chaos_Berthier,shear_Berthier,naraigh2007bubbles,Berti_goodstuff} or of turbulence \cite{Kendon2001,Perlekar2014}, the coarsening can even be arrested~\cite{Ruiz81}.
The aim of this communication is to apply the methodology of these previous works to passive shear-driven phase separation in two and three dimensions, with a view to using the parameters inherent in the shear-flow model as a way of regulating the phase separation. Although current supercomputing platforms enable the simulation of such flows at high resolution in three dimensions~\cite{Perlekar2014, Kendon2001}, it can be noticed that most studies to date on stirred binary mixtures have focused on the two-dimensional case~\cite{Furukawa_scales,Berti_goodstuff,chaos_Berthier,shear_Berthier,naraigh2007bubbles}.  Therefore, a further aim of the present work is to study the effects of dimensionality on the outcome of the phase separation.
The model is the Cahn--Hilliard (CH) equation coupled to an externally-prescribed velocity field: we first of all review this approach before presenting our results.

Phase separation  can be described by the Cahn--Hilliard (CH) equation with constant mobility~\cite{CH_orig}: a concentration field $C(\vecx,t)$  measures the local concentration of the binary liquid, with $C=\pm 1$ denoting saturation in one or other of the components.  Thus, $C=0$ denotes a perfectly mixed state.  It is assumed that the system is in the  spinodal region of the thermodynamic phase space, where the well mixed state is energetically unfavourable.  Consequently, the free energy for the mixture can be modelled as $F[C]=\int \mathd ^d x\left[\tfrac{1}{4}(C^2-1)^2+\tfrac{1}{2}\gamma|\nabla C|^2\right]$,
where the first term promotes  demixing and the second term smooths out sharp gradients in transition zones between demixed regions; also, $\gamma$ is a positive constant, and  $d$ is the dimension of the space.  The twin constraints of mass conservation and energy minimization suggest a gradient-flow dynamics for the evolution of the concentration: $\partial_t C=\nabla \cdot \left[D(C)\nabla \left(\delta F/\delta C\right)\right]$,
where $\delta F/\delta C$ denotes the functional derivative of the free energy and $D(C)\geq 0$ is the mobility function.  In this work, a constant mobility is assumed, such that the basic evolution equation reads
\begin{equation}
\frac{\partial C}{\partial t}=D\nabla^2\left(C^3-C-\gamma\nabla^2C\right).
\label{eq:ch_noflow}
\end{equation}
Models with variable mobility abound~\cite{Bray_LSW,Zhu_numerics,naraigh2007bubbles}, but their characteristics are very similar to the constant-mobility case.  Owing to the simplicity of the latter, a constant mobility is preferred here.

 Equation~\eqref{eq:ch_noflow}  is modified in the presence of an incompressible velocity field $\vecv(\vecx,t)$:
\begin{equation}
\frac{\partial C}{\partial t}+\vecv\cdot\nabla C=D\nabla^2\left(C^3-C-\gamma\nabla^2C\right),\qquad
\nabla\cdot\vecv=0.
\label{eq:ch_flow}
\end{equation}
The velocity  can either be externally prescribed (passive advection) or can arise due to  coupling between  hydrodynamics and phase separation (active advection, modelled using the coupled Navier--Stokes--Cahn--Hilliard equations~\cite{Berti_goodstuff,Kendon2001,Perlekar2014}). The focus of the present work is however on the passive case which can be regarded as an important physical limit of the active case~\cite{shear_Berthier}.
Consideration is given  to symmetric mixtures for which equal amounts of each fluid component are present, such that $\langle C\rangle=0$, where $\langle\cdot\rangle$ denotes spatial averaging.
In view of the flux-conservative nature of the CH equation,  for appropriate boundary conditions on Eqs.~\eqref{eq:ch_noflow}--\eqref{eq:ch_flow}, an initially-symmetric mixture will stay symmetric for all times: $\langle C\rangle(t)=0$, for all $t\geq 0$.
%
%
Since symmetric mixtures exemplify the generic properties of phase separation coupled with flow~\cite{naraigh2007bubbles},  this specialization results in only a little loss of generality.

The dynamics of the unstirred Eq.~\eqref{eq:ch_noflow} admit a constant solution $C_0$, with $C_0=0$ most relevant for symmetric mixtures; linear stability analysis shows this solution to be unstable with respect to small-amplitude perturbations.
This instability is intimately connected to coarsening: starting with the initial condition $C=C_0+\text{[Random fluctuations]}$, local demixing acts  at early times to produce small domains  where $C=\pm 1$.  These domains grow larger over time; more precisely,  the characteristic size $\ell$ of the domains grows as $\ell\sim t^{1/3}$ (\textit{Lifshitz--Slyozov law}).   In the presence of shear flow, the coarsening is interrupted, leading to \textit{coarsening arrest} at a particular length scale set by the flow~\cite{Berti_goodstuff,chaos_Berthier,shear_Berthier,naraigh2007bubbles,Perlekar2014}. 

Certainly, study of the Cahn--Hilliard equation is a well-worn furrow, with much insight available on the analytical side~\cite{Elliott_Zheng}, on the theory of coarsening arrest~\cite{Bray_advphys}, the influence of externally-prescribed flow (References above), and in the area of active mixtures~\cite{Berti_goodstuff}.  The aim of the present paper therefore is to investigate some narrow gaps in the existing literature, in particular the effects of dimensionality on the phase separation under the external shear flow, as well as the effects of flow anisotropy on the outcome of the same.  A further goal in the paper is to establish a number of strict benchmarks that a numerical method must pass in order to be deemed a reliable way of simulating the Cahn--Hilliard dynamics.
%
%
The manuscript is organized as follows. The methodology is first described in Section~\ref{sec:methodology}. The results of some benchmark tests of our code are then presented in Section~\ref{sec:benchmarks}.  Section~\ref{sec:res} contains the results of our investigations concerning coarsening arrest and the effects of dimensionality on the outcome of phase separation.  Section~\ref{sec:res_aniso} concerns the study of the anisotropic shear flow in two and three dimensions in the coarsening arrest regime.  Concluding remarks are presented in Section~\ref{sec:conclusions}.

\section{Methodology} 
\label{sec:methodology}


Equation~\eqref{eq:ch_flow} is solved on a unit cube in $d$ dimensions ($d=2,3$) with periodic boundary conditions in each spatial direction, together with a prescribed initial condition for $C(\vecx,t=0)$.  The equation is discretized in space and time and high-resolution  numerical simulations are used to evolve the initial concentration forward in time.  Two complementary numerical methods are used: a two-dimensional/three-dimensional finite-difference code (FDCH), supplemented by
a lattice-advection model (2DLA).  The 2DLA model is obtained from Ref.~\cite{naraigh2007bubbles} and is revisited briefly here as it provides a way of rapidly generating a large parametric study in two dimensions.  For each method used, the spatial and temporal discretizations are refined until numerical convergence is achieved. 


\paragraph*{Finite-difference code (FDCH):}  In the FDCH code, the concentration field is discretized on a single uniform grid.  All spatial derivatives (even convective terms) are discretized using standard central differences.  For temporal discretization, both the convective term and the nonlinear term $D\nabla^2(C^3-C)$ are treated using the third-order Adams--Bashforth scheme.  The code is stabilized by treating the hyperdiffusion term $-\gamma D \nabla^4 C$ fully implicitly, such that at each timestep, the following linear problem must be solved:
\begin{eqnarray}
(1 & + & \gamma D \Delta t \nabla^4)C^{n+1}_{ijk} \nonumber \\
& = & C^n_{ijk}+\Delta t\left[\tfrac{23}{12}\mathcal{C}^n_{ijk}-\tfrac{16}{12}\mathcal{C}^{n-1}_{ijk}+\tfrac{5}{12}\mathcal{C}^{n-2}_{ijk}\right] \nonumber \\
& := & \mathrm{RHS},
\label{eq:numerics1}
\end{eqnarray}
where $\mathcal{C}=-\vecv\cdot\nabla^2 C+D\nabla^2(C^3-C)$, the superscript $n$ denotes the temporal discretization, and the subscript $(i,j,k)$ denotes a position on the discrete spatial grid.

To determine $C^{n+1}_{ijk}$ and hence to evolve the state of the system forward in time, Eq.~\eqref{eq:numerics1} requires the inversion of a fourth-order operator.   Because of the parallel-computing method used (discussed below), it is more computationally efficient to solve a Helmholtz-like problem:
\begin{equation}
\left(1-\alpha \nabla^2\right)^2C^{n+1}_{ijk}=\mathrm{RHS}-2\alpha\nabla^2 C^{n+1}_{ijk},\qquad \alpha=\sqrt{\gamma D \Delta t}.
\label{eq:numerics2}
\end{equation}
By reducing the problem~\eqref{eq:numerics1} to this more canonical form, a large price is paid: the right-hand side of Eq.~\eqref{eq:numerics2} is indeterminate.  However, in practice the term $C^{n+1}_{ijk}$ can be replaced by an interpolation of $C_{ijk}$ obtained from prior times, and optimal interpolation coefficients can be chosen.  The expression
\begin{equation}
\left(1-\alpha \nabla^2\right)^2C^{n+1}_{ijk}=\mathrm{RHS}-2\alpha\nabla^2 \left[\tfrac{3}{2}C^{n}_{ijk}-\tfrac{1}{2}C^{n-1}_{ijk}\right]
\label{eq:numerics3}
\end{equation}
yields a numerical solution that is in excellent agreement with a number of extremely precise benchmarks (See Section~\ref{sec:benchmarks}).  Equation~\eqref{eq:numerics3} can therefore be inverted by solving a pair of Helmholtz equations:
\begin{eqnarray*}
\left(1-\alpha \nabla^2\right)C^{n+1}_{ijk}&=&\omega_{ijk},\\
\left(1-\alpha \nabla^2\right)\omega^{n+1}_{ijk}&=&\mathrm{RHS}-2\alpha\nabla^2 \left[\tfrac{3}{2}C^{n}_{ijk}-\tfrac{1}{2}C^{n-1}_{ijk}\right].
\end{eqnarray*}
In practice, this equation pair was solved using standard successive over-relaxation methods.
Finally, the numerical method was coded in Fortran 90.  The method was parallelized by domain decomposition using MPI, with a domain-decomposition scheme that was the same in all three spatial dimensions.  Validation tests of the FDCH code are  presented in Sec.~\ref{sec:benchmarks}.


%
%

\paragraph*{Lattice advection model (2DLA):}  The 2DLA code is an operating-splitting technique wherein an advection half-step is performed first, followed by a Cahn--Hilliard diffusion half-step.  The Cahn--Hilliard equation is discretized on to a uniform grid (the `lattice').   The model flows suitable for this method have no stagnation points, such that at each timestep, the lattice is mapped to itself under a bijective map defined by the flow field.  Thus, the advection half-step amounts to a Lagrangian scheme with a natural interpolation on to the underlying Eulerian grid (the lattice).  Consequently, the combined half-steps are a hybrid
Eulerian-Lagrangian method, and the timestep can be very large (no CFL condition).  In particular, for the model flow in Section~\ref{sec:res} the timestep can be as large as the flow quasi-period in the case of small Cahn--Hilliard diffusivities.  The details of the 2DLA code are presented elsewhere, together with extensive validation of the method~\cite{naraigh2007bubbles}.

\section{Benchmark tests of the numerical method FDCH} 
\label{sec:benchmarks}

In this section the Cahn--Hilliard equation without flow is solved numerically, using the FDCH, in three distinct contexts.  The aim is not only to validate the FDCH numerical method described in Sec.~\ref{sec:methodology}, but also to describe more generally three benchmark problems that can be used to stress-test any numerical method for the Cahn--Hilliard equation.

\subsection{Linear stability analysis}

The dynamics of the unstirred Eq.~\eqref{eq:ch_noflow} admit a constant solution $C_0$.  Because of our focus here on symmetric mixtures it suffices to consider the constant solution $C_0=0$.  This is linearly unstable: by inserting the trial solution $C=C_0+\delta C(\vecx,t)$ into Eq.~\eqref{eq:ch_noflow} and omitting nonlinear terms in $\delta C$, the equation $(\partial/\partial t)\delta C=-D\nabla^2\delta C-\gamma D\nabla^4\delta C$ is obtained, with normal-mode solution $\delta C\propto \mathe^{\sigma t+\imag\veck\cdot\vecx}$, where $\veck$ is the wave vector and $\sigma$ is the growth rate, connected to the wave vector through the dispersion relation
\begin{equation}
\sigma(\veck)=D (k^2-\gamma  k^4),\qquad k=|\veck|.
\label{eq:linear_dispersion}
\end{equation}
Thus, the base state $C=C_0=0$ is always unstable.  The cutoff wavenumber is $k_\mathrm{c}=\gamma^{-1/2}$ and the most-dangerous mode is $k_{\mathrm{max}}=k_\mathrm{c}/\sqrt{2}$.

\begin{figure}[htb]
\centering
\includegraphics[width=0.5\textwidth]{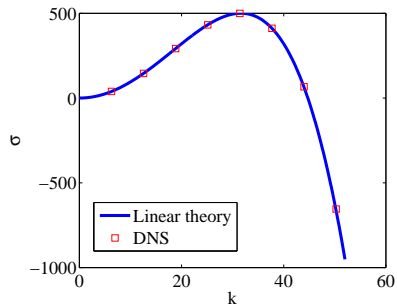}
\caption{Comparison between linear theory and direct numerical simulation.  Model parameters: $D=1$ and $\gamma=5\times 10^{-4}$.  Simulation parameters: $\Delta x=1/304$, $\Delta t=10^{-6}$.}
\label{fig:compare_linear}
\end{figure}

The FDCH code is configured as one-dimensional model by fixing the number of gridpoints in the $y$ and $z$ directions to be 1, such that only the $x$-direction is relevant.  The initial condition $C(x,t=0)=\epsilon\cos(n(2\pi/L)x)$ is prescribed, where the size of the domain in the $x$-direction is $L=1$, where $\epsilon=10^{-4}$, and where $n$ is a positive integer.  The code is executed for a range of $n$-values.  For each $n$-value, a time series $\|C\|_2(t)$ is obtained, where $\|C\|_2=\left[\int \mathd x\, C(x,t)^2\right]^{1/2}$ is the $L^2$ norm of the concentration.  The quantity $\|C\|_2(t)$ is seen to grow exponentially at the growth rate $\sigma(k=n(2\pi/L))$, consistent with the linear theory in Eq.~\eqref{eq:linear_dispersion}.  A comparison between the linear theory in Eq.~\eqref{eq:linear_dispersion} and the numerical growth rates is shown in Fig.~\ref{fig:compare_linear}, and excellent agreement is obtained.

\subsection{Weakly nonlinear analysis}

The linear theory exemplified by Eq.~\eqref{eq:linear_dispersion} represents a dramatic simplification of the original problem~\eqref{eq:ch_noflow}, and is obviously limited in its applicability to short times where the quantity $\|C\|_2$ remains small, such that nonlinear terms in Eq.~\eqref{eq:ch_noflow} are negligible.  However, (again in one spatial dimension) the applicability of the theory can be extended in a \textit{barely supercritical} regime wherein only a single unstable mode fits inside the periodic domain $[0,L]$.  Parameters appropriate for this regime are $L=1$ and $\gamma=1/8\pi^2$.  The most-dangerous mode therefore occurs at $k_1=2\pi/L$ and the cutoff is at $k_\mathrm{c}=\sqrt{2}k_1$, with $k_1<k_\mathrm{c}<2k_1$, such that precisely one unstable mode fits inside the box.

The complete solution of Eq.~\eqref{eq:ch_noflow} is expanded as a Fourier series,
\begin{equation}
C(x,t)=\sum_{n=-\infty}^\infty A_n(t)\mathe^{\imag k_n x},\qquad 
k_n=(2\pi/L)n,\qquad A_{-n}=A_n^*,
\label{eq:fourier}
\end{equation}
with $A_0=0$ for symmetric mixtures.  The solution~\eqref{eq:fourier} is substituted into Eq.~\eqref{eq:ch_noflow}.  One obtains the following amplitude equations:
\begin{equation}
\frac{dA_n}{dt}=\sigma(k_n)A_n-Dk_n^2\sum_{p=-\infty}^\infty \sum_{q=-\infty}^\infty A_p A_q A_{n-p-q}.
\label{eq:fourier1}
\end{equation}
For a barely-supercritical system, the fundamental mode $(n=\pm 1)$ has a positive linear growth rate, while all other modes have a negative linear growth rate.  Initially therefore, the fundamental dominates the evolution.  Overtones will only be relevant if they couple to the fundamental.  We therefore simplify Eq.~\eqref{eq:fourier1} by considering the dominant modes.  These will be the fundamental and a handful of overtones.  In view of the cubic nature of the nonlinearity, an initial condition containing only the fundamental will evolve into a disturbance containing only odd multiples of the fundamental.  We therefore reduce the equations down to a triple by considering the fundamental and the $n=3,5$ overtones, and by neglecting all other modes.  A further simplification occurs in the overtone equations, wherein one considers the most-dominant interaction terms only; \textit{i.e.} those that are quadratic in $A_1$.  Thus, we arrive at the following set of equations
\begin{subequations}
\begin{eqnarray}
\frac{dA_1}{dt}&=&\sigma(k_1)A_1-Dk_1^2A_1\left(6|A_5|^2+6|A_3|^2+4|A_1|^2\right) \nonumber\\
& - & 3Dk_1^2A_3\left(A_3A_5^*+A_1^*A_1^*\right)-6A_1^*A_3^*A_5, \label{eq:sl0}\\
\frac{dA_3}{dt}&=&\sigma(k_3)A_3-Dk_3^2A_1^3,\\
\frac{dA_5}{dt}&=&\sigma(k_5)A_5-2Dk_5^2A_1^2A_3.
\end{eqnarray}%
\end{subequations}%
The slow-manifold approximation is made~\cite{pavliotis2008multiscale}, whereby the left-hand side of the overtone equations is set to zero, giving
\begin{equation}
A_3=\frac{9Dk_1^2}{\sigma(k_3)}A_1^3,\qquad A_5=\frac{450D^2k_1^4}{\sigma(k_3)\sigma(k_5)}A_1^5.
\label{eq:sl1}
\end{equation}
Of crucial relevance here is the fact that $A_n\propto A_1^n$ valid at least for $n=3,5$.  This is a particular case of the celebrated Stuart--Landau theory~\cite{schmid2001stability}.  The Stuart--Landau law~\eqref{eq:sl1} is substituted back into Eq.~\eqref{eq:sl0}.  One obtains a nonlinear evolution equation for the fundamental:
\begin{equation}
\frac{dA_1}{dt}=\sigma(k_1)A_1- Dk_1^2 A_1 P(|A_1|^2),
\label{eq:sl2}
\end{equation}
where $P(\cdot)$ is a fifth-order polynomial.  For the relevant case $\gamma=1/8\pi^2$, the coefficients of $P(x)$ are readily computed explicitly as rational numbers (\textit{e.g.} using a symbolic algebra package), and again using a symbolic algebra package, it is readily shown that $P(x)$ has no real roots, and moreover, that $P(x)>0$.  The nonlinearity in Eq.~\eqref{eq:sl2} is therefore saturating.
%
%
%
%

\begin{figure}[htb]
\centering
\includegraphics[width=0.6\textwidth]{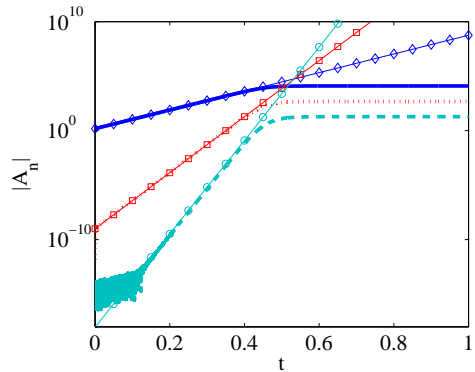}
\caption{Comparison between weakly linear theory and direct numerical simulation.  Model parameters: $D=1$, $L=1$, and $\gamma=1/8\pi^2$.  Simulation parameters: $\Delta x=1/304$, $\Delta t=10^{-8}$.  The simulation results are shown in thick lines: solid line: $|A_1|$, dotted line: $|A_3|$, dashed line: $|A_5|$.  The predictions from weakly nonlinear theory are shown in thin lines with symbols.  Diamonds: $\mathe^{\sigma(k_1) t}$, squares: $\mathe^{3\sigma(k_1) t}$, circles: $\mathe^{5\sigma(k_1) t}$.}
\label{fig:wnl_result}
\end{figure}

The foregoing theory was compared to the results of a direct numerical simulation seeded with the initial condition $C_\mathrm{init}=\epsilon \cos(k_1 x)$, with $\epsilon=10^{-4}$.  A spectral analysis of the numerical solution was obtained and the results plotted in Fig.~\ref{fig:wnl_result}.  Excellent agreement is obtained, thereby confirming the validity of the FDCH code.

\subsection{Lifshitz--Slyozov scaling law}

A final benchmark case concerns the simulation of phase separation without flow in three dimensions.  The Cahn--Hilliard equation~\eqref{eq:ch_noflow} was simulated using the FDCH on a unit cube with periodic boundary conditions in each dimension.  The initial condition  was random, with $C_\mathrm{init}(\vecx)$ assigned a different random value (drawn from the uniform distribution) in the range $[-0.1,0.1]$ at each point $\vecx$.
 Three distinct spatial resolutions were investigated: $313^3$ (`low'), $505^3$ (`medium'), and $707^3$ (`high').  The high-resolution case corresponds to a simulation with over 300 million gridpoints. The model parameters are $D=L=1$ and $\gamma=10^{-5}$, corresponding to a  broad spectrum of linearly unstable modes.  A timestep $\Delta t=10^{-5}$ was used in each case. 
%

\begin{figure}[t]
	\centering
	\subfigure[$\,\,t=0.002$]{\includegraphics[width=0.236\textwidth]{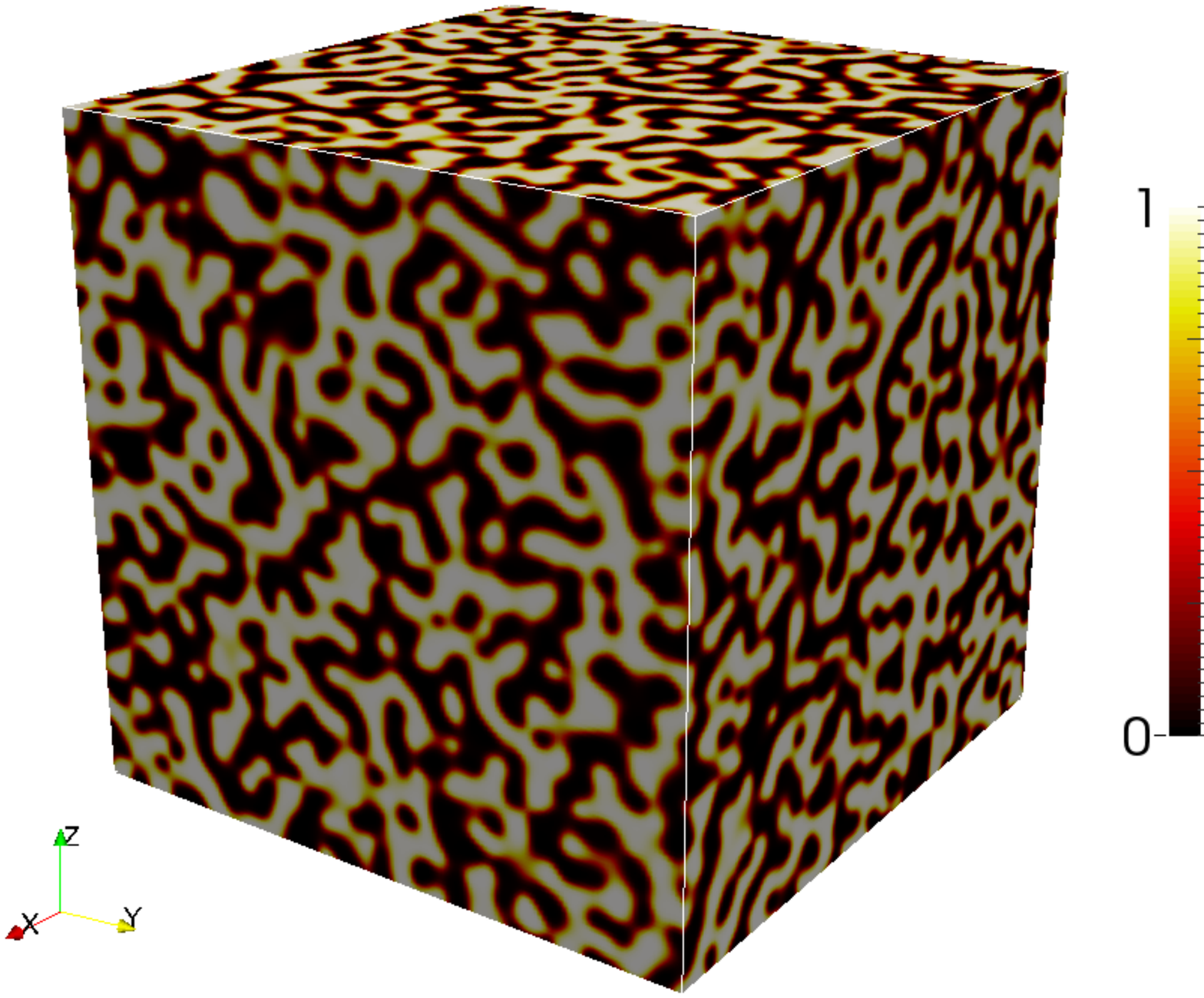}}
	\subfigure[$\,\,t=0.006$]{\includegraphics[width=0.2\textwidth]{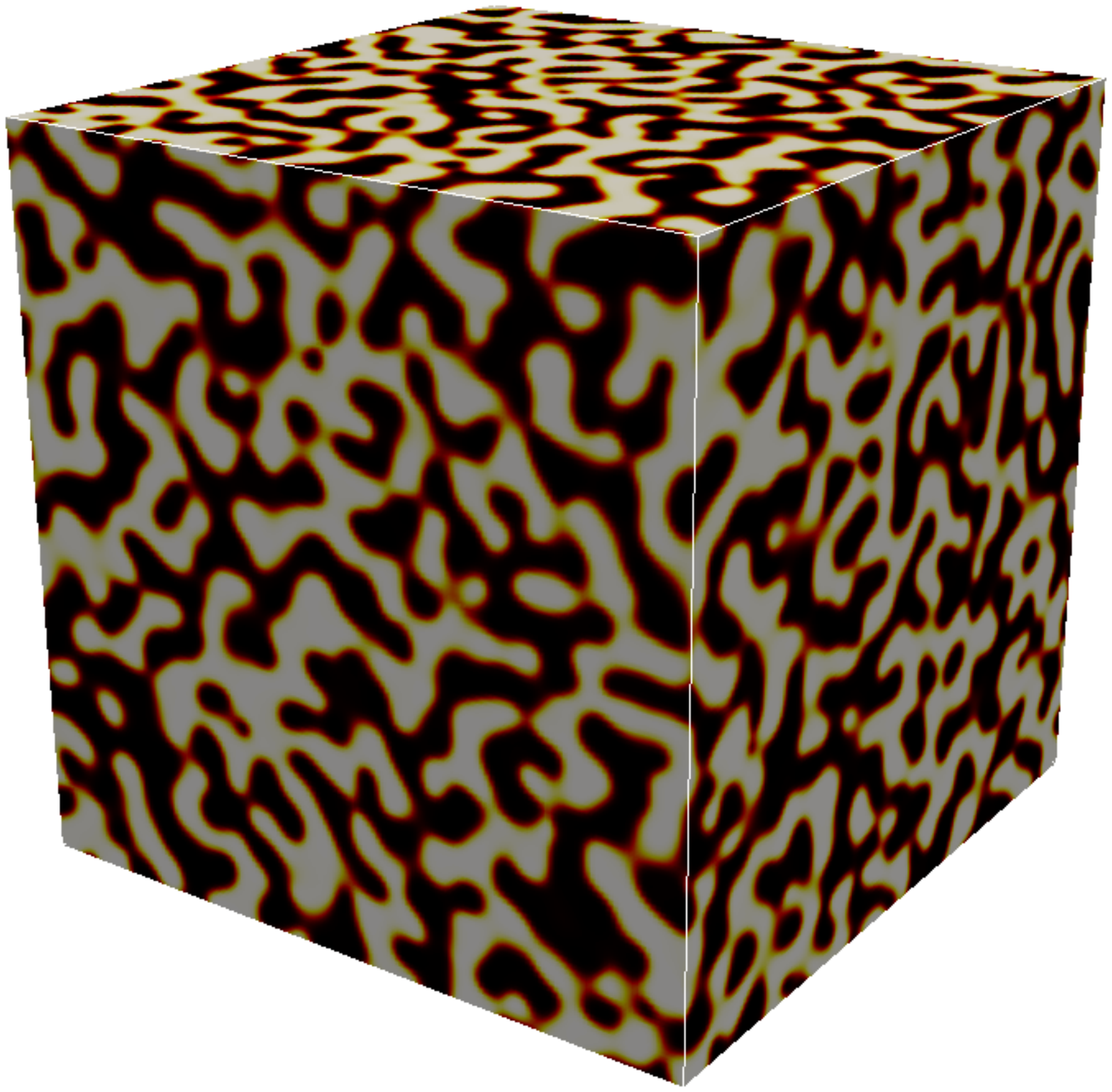}}
	\subfigure[$\,\,t=0.03$]{\includegraphics[width=0.209\textwidth]{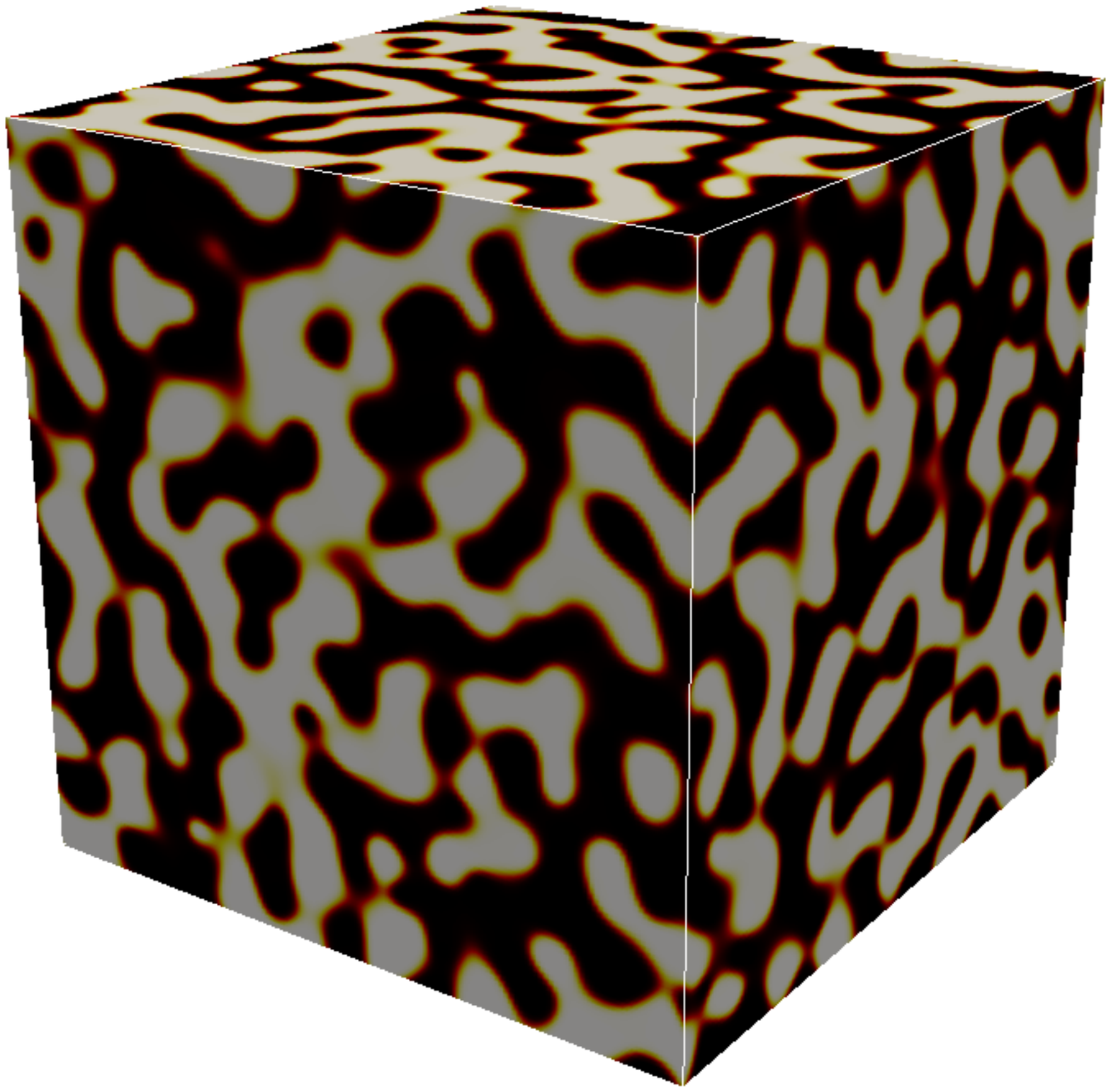}}
	\subfigure[$\,\,t=0.16$]{\includegraphics[width=0.19\textwidth]{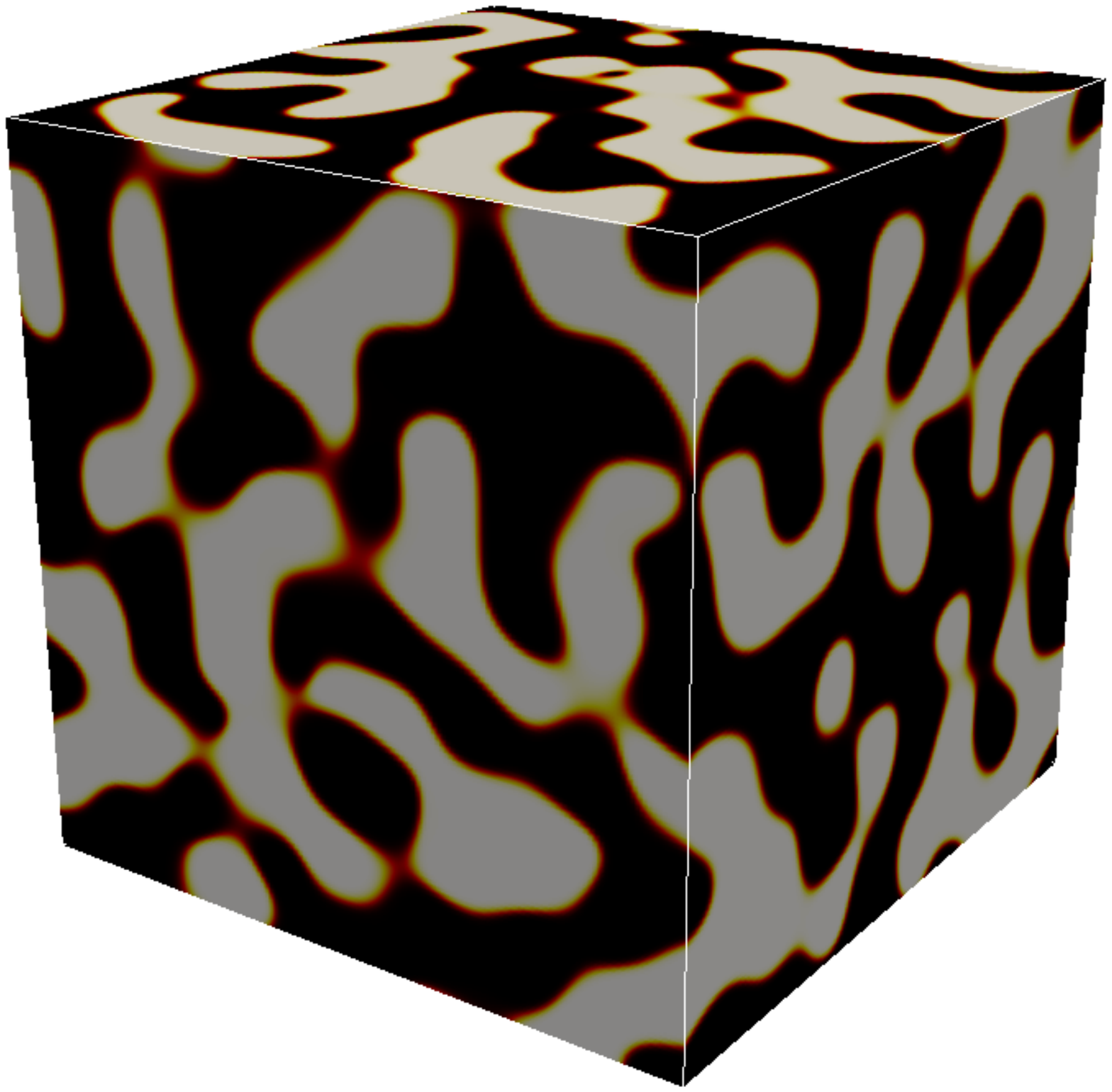}}
	\caption{Snapshots of phase separation for the Cahn--Hilliard equation without flow (medium resolution).  Shown is an isovolume plot of $\phi=(C+1)/2$, with $\phi$ ranging from $0$ to $1$.  The initial condition and the simulation parameters are given in the text.}
	\label{fig:snapshot}
\end{figure}

\begin{figure}[htb]
	\centering
		\includegraphics[width=0.6\textwidth]{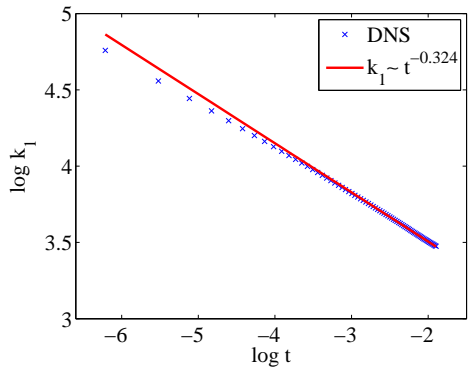}
		\caption{Results for the medium-scale simulation.  Time dependence of $k_1$, with fitted power-law exponent $0.324$ close to the theoretical Lifshitz--Slyozov exponent $1/3$.}
	\label{fig:fig_scaling_no_flow}
\end{figure}

Snapshots of the three-dimensional concentration field at various times are shown in Fig.~\ref{fig:snapshot} for the medium-resolution case.  These figures exhibit clearly the phenomenon of domain coarsening.
To determine the coarsening dynamics precisely, the typical domain size was computed as $\ell=1/k_1$, where $k_1$ is a typical wavenumber obtained from structure-function calculations (Appendix~\ref{app:ellsf}).  Calculations based on $\ell\propto (1-\langle C^2\rangle)^{-1}$ yielded very similar results.
The results of the foregoing analysis are presented in Fig.~\ref{fig:fig_scaling_no_flow}.  The results were found to be the same for the `small', 'medium' and `large' runs.  Moreover, a further run with the `small' spatial resolution wherein the timestep was halved also gave the same results.  Thus, it suffices to present a representative sample result, and the scaling law for the medium-scale run is therefore presented in the figure.
Based on Fig.~\ref{fig:fig_scaling_no_flow}, a power law is justified at late times (we take $t\geq 0.08$).  This is consistent with the fact that the Lifshitz--Slyozov scaling law is merely \textit{asymptotic}, and that a more complicated dynamics pertains at finite times~\cite{Huse1986}.  Again based on Fig.~\ref{fig:fig_scaling_no_flow}, a power-law fit
$\ell \sim t^{0.324}$ is reported, close to the predicted scaling behaviour $\ell \sim t^{1/3}$, based on Lifshitz--Slyozov theory.  A possible reason for the small discrepancy is the presence of finite-size effects at late time, which can further spoil the scaling laws.  Nevertheless, the fitted exponent $0.324$ is very close to the fitted exponent reported in the literature for the two-dimensional case, the latter obtained using a method independent of the present approach and at very high accuracy and spatial resolutions (Reference~\cite{naraigh2007bubbles} reports a fitted exponent $0.325$ at spatial resolution $512^2$, using pseudo-spectral methods).  Thus, we are satisfied not only with the correctness of the code, but also with the similarity of the coarsening dynamics (without flow) in two and three dimensions.
\section{Results - Isotropic flows} 
\label{sec:res}


The effect of chaotic shear flow via passive advection is modelled with a random-phase sine flow with quasi-period $\tau$ such that, in the 3D case, at time $t$,
\begin{subequations}
\begin{eqnarray}
u&=&A\sin\left[k_0 (y+z)+\varphi\right],\qquad  0\leq \text{mod}(t,\tau)<\tfrac{1}{3}\tau,\\
v&=&A\sin\left[k_0 (x+z)+\psi\right],\qquad     \tfrac{1}{3}\tau\leq \text{mod}(t,\tau)<\tfrac{2}{3}\tau,\\
w&=&A\sin\left[k_0 (x+y)+\chi\right],\qquad     \tfrac{2}{3}\tau\leq \text{mod}(t,\tau)<\tau,
\end{eqnarray}%
\label{eq:flow_def}%
\end{subequations}%
where the velocity components not listed are zero and where $t$ is written uniquely as $t=\myq\tau+\mu$ for $\myq$ zero or a positive integer, with $0\leq \mu<\tau$, and hence $\mathrm{mod}(t,\tau):=\mu$.  The random phases $(\varphi,\psi,\chi)$ are renewed after $N$ quasi-periods $\tau$.  The flow therefore has two timescales, with $\tau_{\mathrm{corr}}=N\tau$ being the correlation time. 
Correspondingly, the 2D simulations are obtained using a flow analogous to Eq.~\eqref{eq:flow_def}:
\begin{subequations}
\begin{eqnarray}
u&=&A\sin\left(k_0 y+\varphi\right),\qquad   0\leq \text{mod}(t,\tau)<\tfrac{1}{2}\tau,\\
v&=&A\sin\left(k_0 x+\psi\right),\qquad      \tfrac{1}{2}\tau\leq \text{mod}(t,\tau)<\tau,
\end{eqnarray}%
\label{eq:flow_def_2d}%
\end{subequations}%
where the velocity components not listed are zero and all other symbols have the same meaning as in Eq.~\eqref{eq:flow_def}.
Based on the velocity fields~\eqref{eq:flow_def}--\eqref{eq:flow_def_2d}, Eq.~\eqref{eq:ch_flow} is solved in units for which the mean velocity 
\begin{equation}
U=\sqrt{2}\left(\lim_{T\rightarrow\infty}T^{-1}\int_0^T \langle \vecv^2\rangle\,\mathd t\right)^{1/2}
\label{eq:udef}
\end{equation}
and box size $L$ are both unity (hence, $A=1$ in Equations~\eqref{eq:flow_def}--\eqref{eq:flow_def_2d} also), meaning that there are three independent flow parameters $(\tau,N,k_0)$.  The corresponding nondimensional diffusion parameters are $\mydiff=D/UL$ and $\gammanondim=\gamma/L^2$, which correspond respectively to the inverse P\'eclet number and the square of the Cahn number. 
%
%

\begin{figure}[htb]
	\centering
		\includegraphics[width=0.5\textwidth]{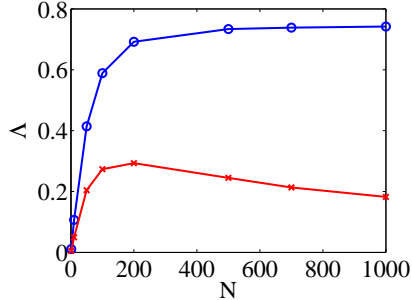}
		\caption{(Circles) Average value of the maximal fully-converged infinite-time Lyapunov exponent $\Lambda(N,k_0,\tau)$ of the flow~\eqref{eq:flow_def}--\eqref{eq:flow_def_2d}, with  $\tau=0.01$ and $k_0=2\pi$ (box size $L=1$).  Circles: 3D case; Crosses: 2D case.}
	\label{fig:mylep}
\end{figure}

A preliminary investigation here concerns the derivation of a measure of the mean strain rate associated with the flows~\eqref{eq:flow_def}--\eqref{eq:flow_def_2d}, to be denoted by $\mylep$.
Not only must  $\mylep$ take account of the flow amplitude and flow lengthscale, but also the flow timescales.  Thus, the mean rate of strain is identified with the average value of the maximal Lyapunov exponent of the flow, computed in a standard fashion in both two and three dimensions~\cite{naraighPhD}. The results are shown in Fig.~\ref{fig:mylep}.  Care is needed in interpreting the results reported in this figure: for each parameter set $(N,k_0,\tau)$ a finite time Lyapunov exponent $\Lambda_{\myq}(\vecx_0)$ is calculated for a trajectory starting at $\vecx_0$ and evolved forwards in time through $\myq$ iterations of the sine-flow map (which in turn is based on Equation~\eqref{eq:flow_def} or~\eqref{eq:flow_def_2d}).  The result is then averaged over a large ensemble of initial points $\vecx_0$ and an average value $\langle \Lambda_{\myq}\rangle$ is obtained.  The averaged result is then investigated for convergence in the limit as $\myq\rightarrow\infty$.  Only when convergence is achieved is the infinite-time Lyapunov exponent $\Lambda(N,k_0,\tau)$ identified, the results of which are shown in Fig.~\ref{fig:mylep}.

A campaign of simulations based on Eq.~\eqref{eq:ch_flow} in the presence of the shear flows (\ref{eq:flow_def},\ref{eq:flow_def_2d}) is carried out, the details of which are summarized in Tab.~\ref{tab:simsx}.  
\begin{table}
\begin{center}
		\begin{tabular}{|c|c|c|c|}
		\hline
			  & Grid size & $\tau_{\mathrm{corr}}$ & $\gammanondim$ \\
			\hline
		  \hline	
			2DLA & $512^2$ & Various & $1.5\times 10^{-5}$  \\
			FDCH2D-Long & $314^2$ & 1  & $10^{-4}$ \\
			FDCH2D-Short &   $314^2$ & 0.01 & $10^{-4}$ \\
			FDCH3D-Long &    $314^3$ & 1 & $10^{-4}$  \\
			FDCH3D-Short &   $314^3$ &  0.01 & $10^{-4}$ \\
			\hline
		\end{tabular}
\caption{Summary of simulations performed in Section~\ref{sec:res}.  Here $\tau_\mathrm{corr}=N\tau$, where $\tau$ is the quasi-period and $N$ is the number of quasi-periods after which the random phases of the chaotic flow are renewed.  Also, $\mathcal{C}_\mathrm{n}$ is the square of the Cahn number.}
\label{tab:simsx}
\end{center}
\end{table}
In the FDCH code, the flow quasi-period is set to $\tau=0.01$, such that the correlation time  is $\tau_{\mathrm{corr}}=1$ for the `FDCH-Long' simulations and $\tau_\mathrm{corr}=0.01$ for the `FDCH-Short' simulations (recall, $\tau_{\mathrm{corr}}=N\tau$).   For the 2DLA code, a range of flow correlation times is possible, relevant values of which are alluded to in the text.
%
%
Also in the 2DLA simulations, a lattice method is used, such that the timestep can be set equal to the flow quasi-period (as discussed in Sec.~\ref{sec:methodology}).  For all other simulations the timestep is $\Delta t=10^{-4}$ or $\Delta t=10^{-5}$ depending on the requirements for the numerical method to converge.

A brief review of the results in Table~\ref{tab:simsx} reveals that coarsening arrest occurs in all simulations.  For certain cases drawn from the simulation runs 2DLA, FDCH2D-Long and FDCH3D-Long,  the variance $\myvar(t)=\langle C^2\rangle^{1/2}$ decays exponentially.  The full details are presented in the subsections below.
The focus in subsections~A, B  is on the results of the study of passive chaotic shear-driven phase separation, reported for the first time -- to our knowledge -- in three dimensions.   However, these results echo the findings of previous studies in two dimensions.  Therefore, a key point of departure is subsection C  wherein the differences between the 2D and 3D cases are made manifest in the probability distribution function of the concentration field in the hyperdiffusive regime.

%


\subsection{Characterization of the different regimes} 
\label{subsec:res_gen}

\begin{figure}
	\centering
	\subfigure[$\,\,t=1$]{\includegraphics[width=0.32\textwidth]{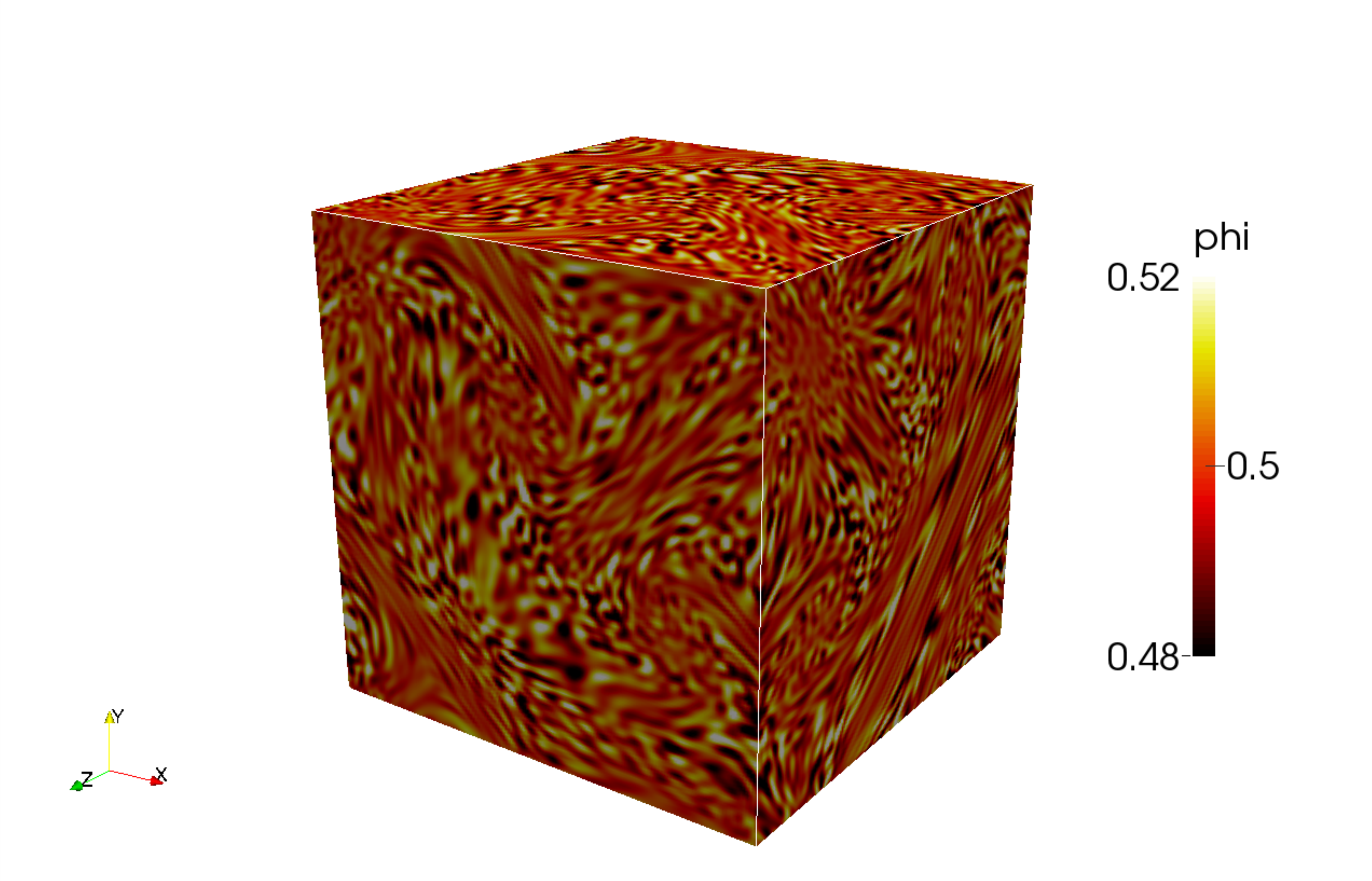}}
	\hspace{-.2in}
	\subfigure[$\,\,t=4$]{\includegraphics[width=0.32\textwidth]{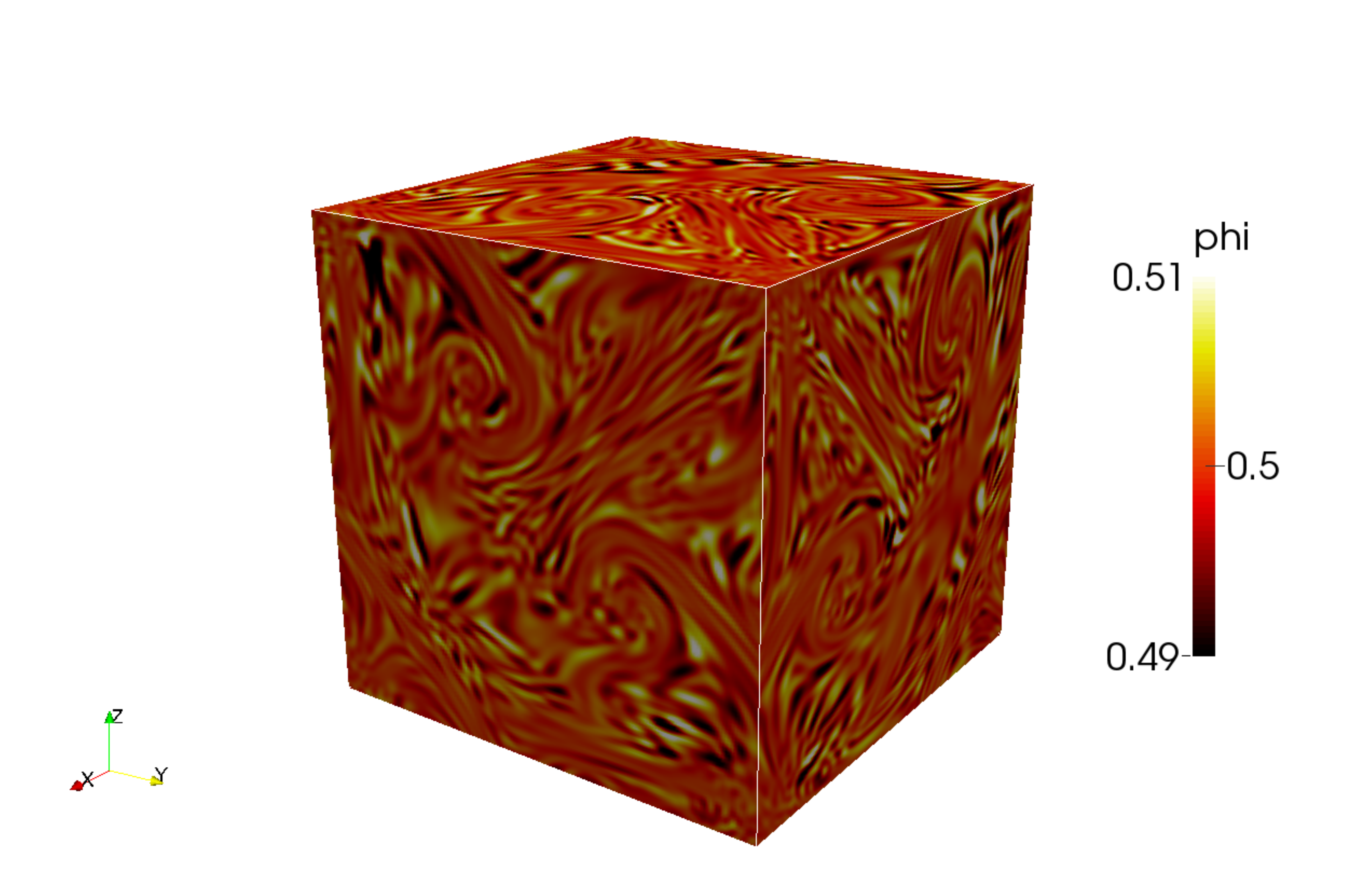}}
	\hspace{-.2in}
	\subfigure[$\,\,t=8$]{\includegraphics[width=0.32\textwidth]{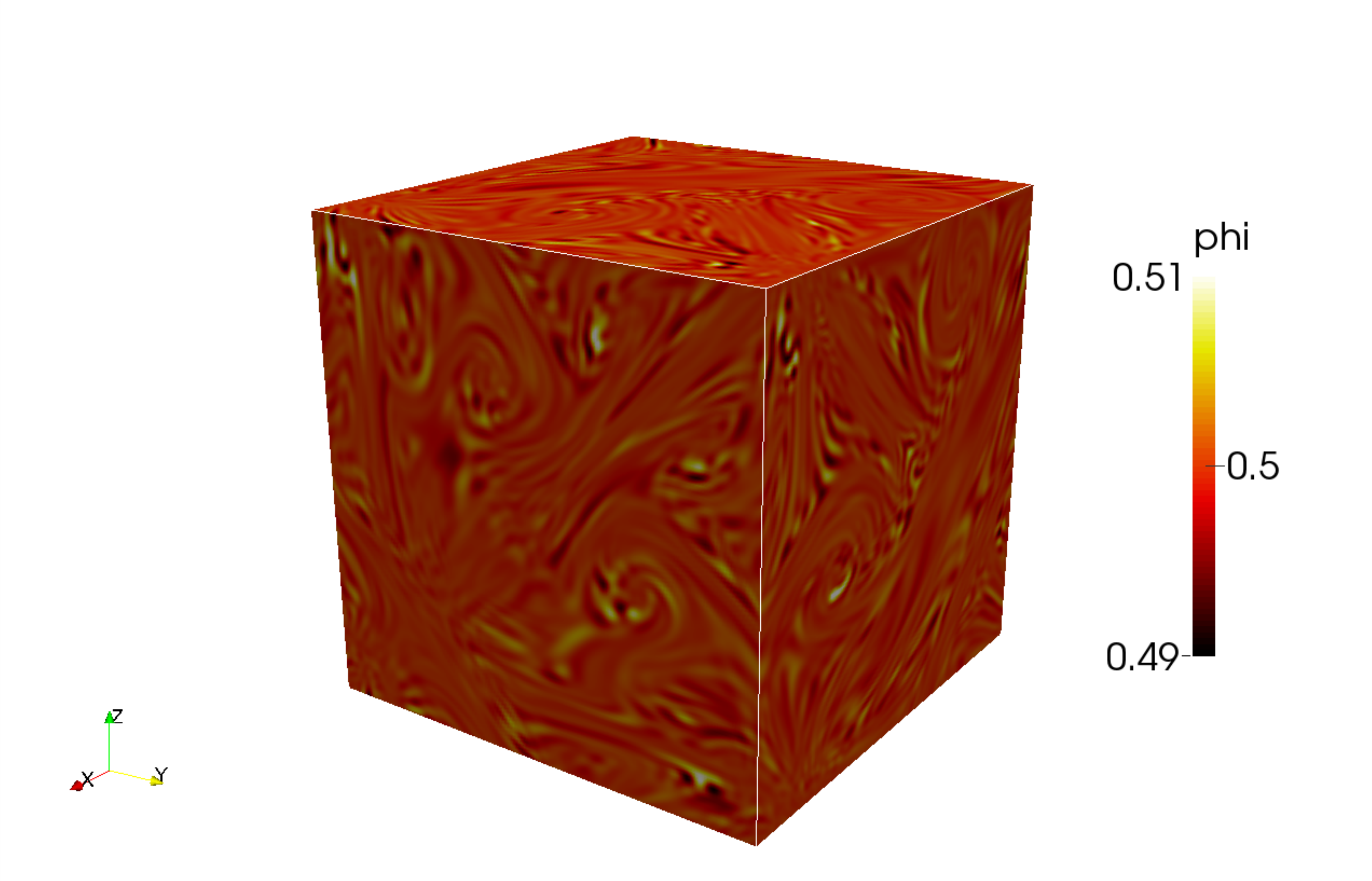}}\\
	\subfigure[$\,\,t=1$]{\includegraphics[width=0.32\textwidth]{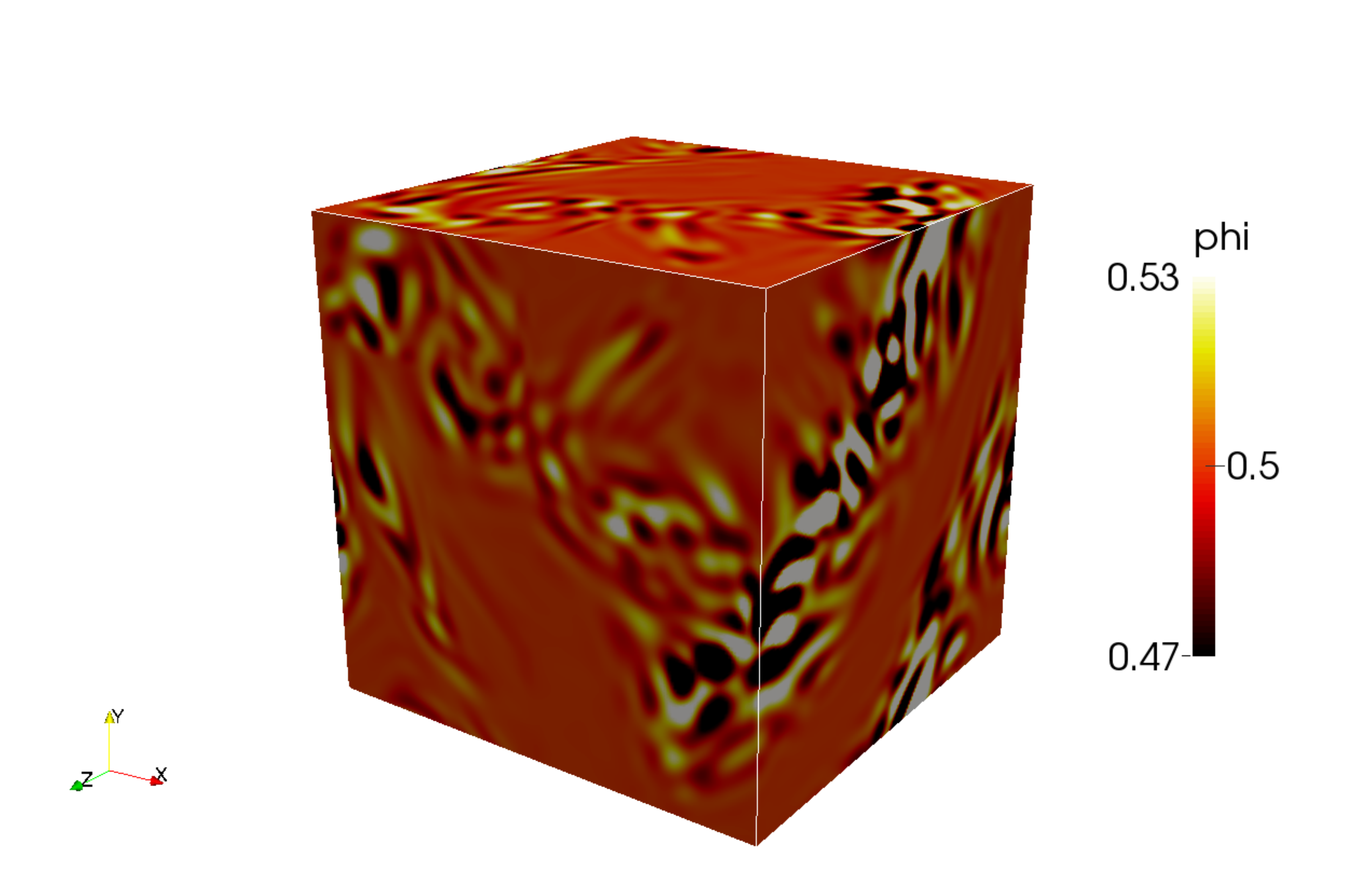}}
	\hspace{-.2in}
	\subfigure[$\,\,t=4$]{\includegraphics[width=0.32\textwidth]{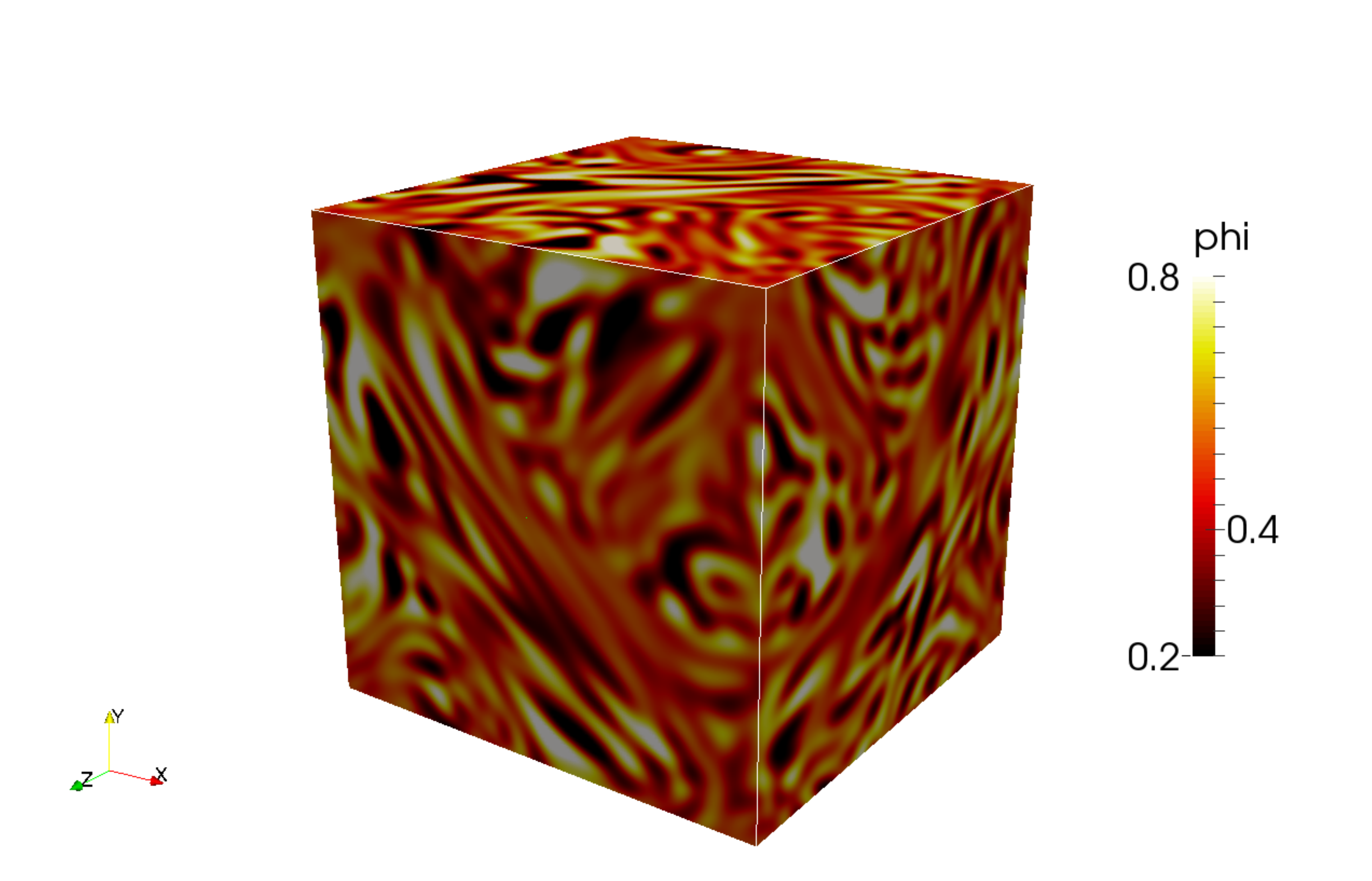}}
	\hspace{-.2in}
	\subfigure[$\,\,t=8$]{\includegraphics[width=0.32\textwidth]{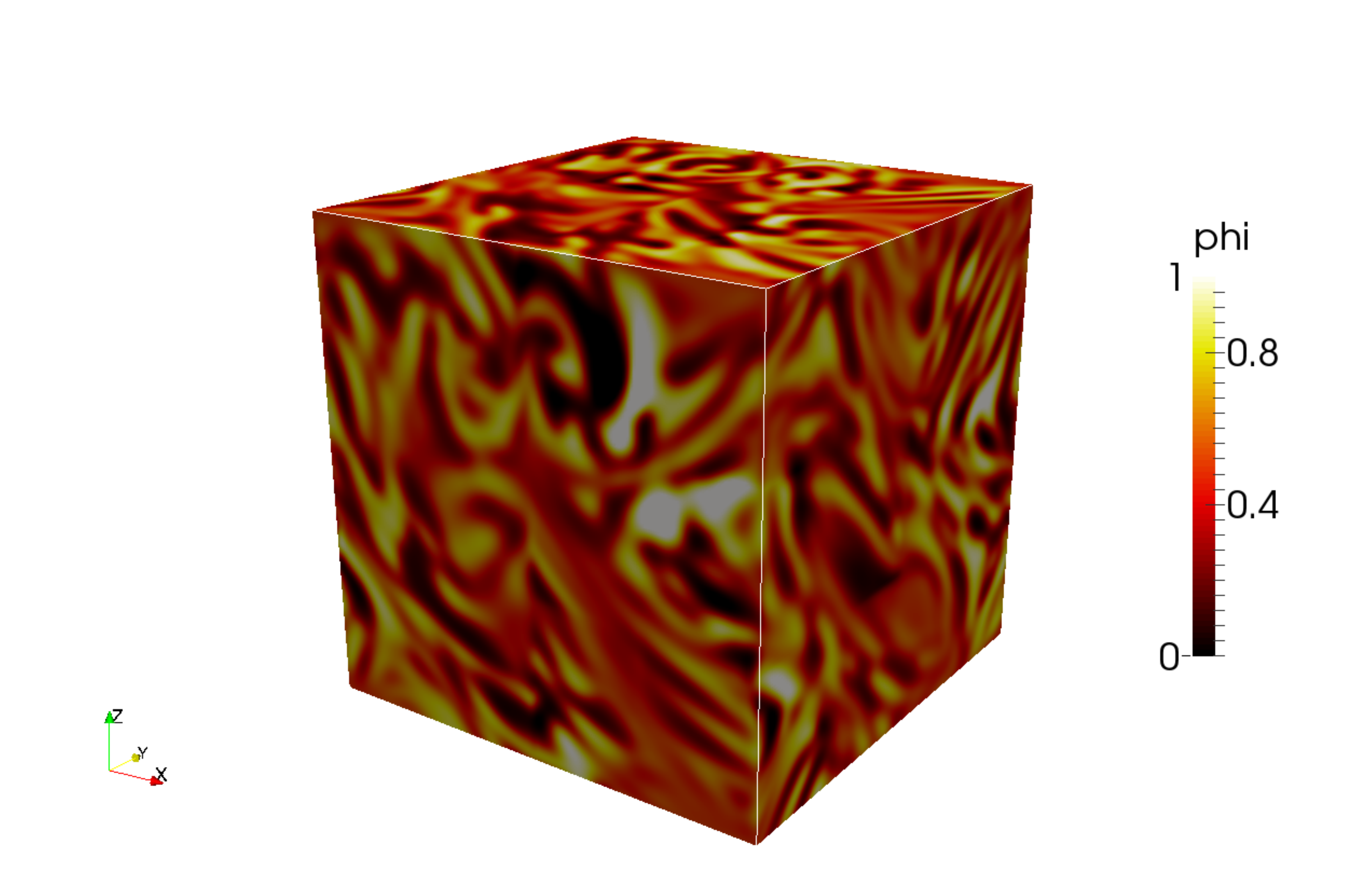}}\\
		\subfigure[$\,\,t=1$]{\includegraphics[width=0.32\textwidth]{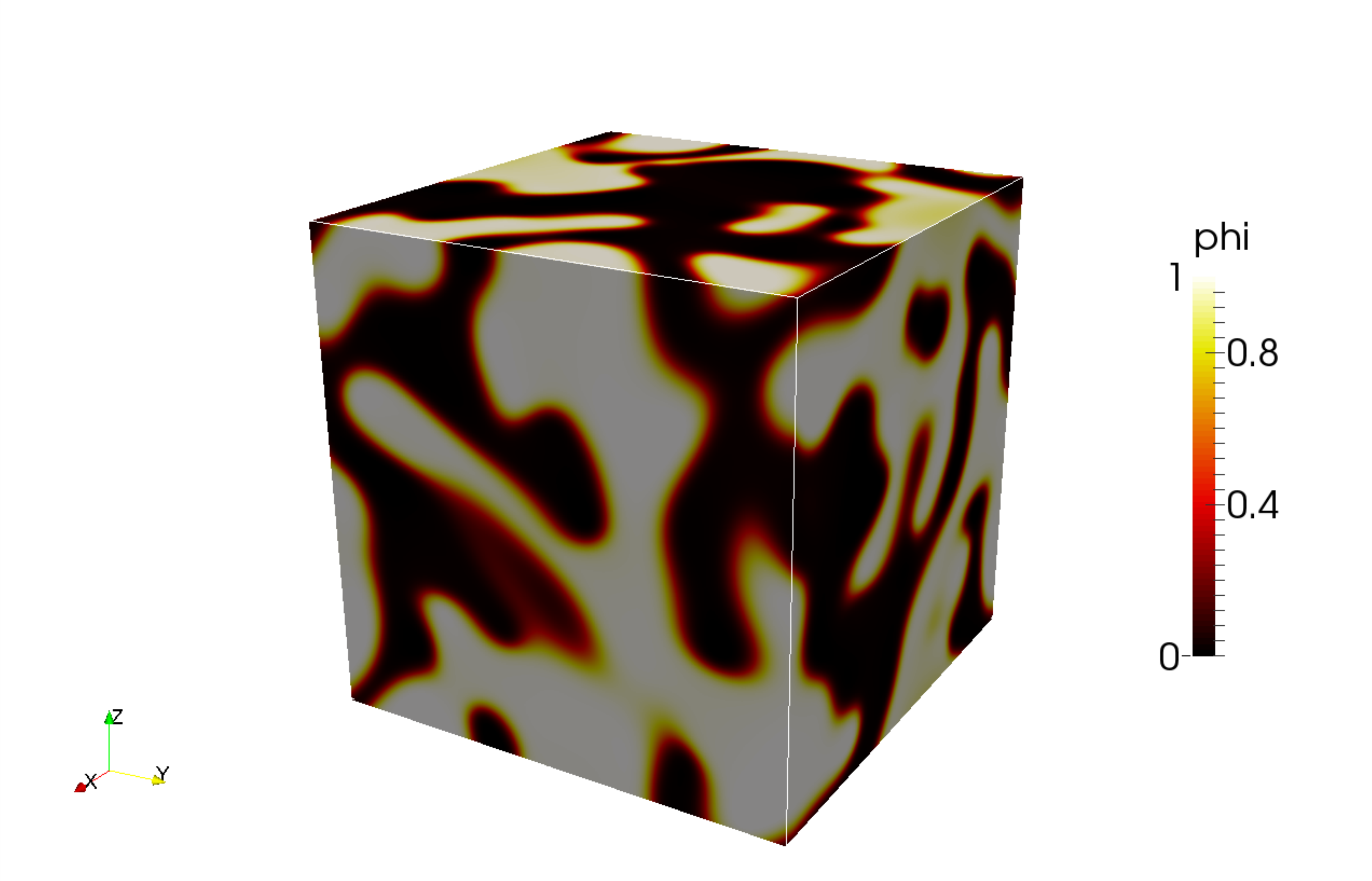}}
		\hspace{-.2in}
	\subfigure[$\,\,t=2$]{\includegraphics[width=0.32\textwidth]{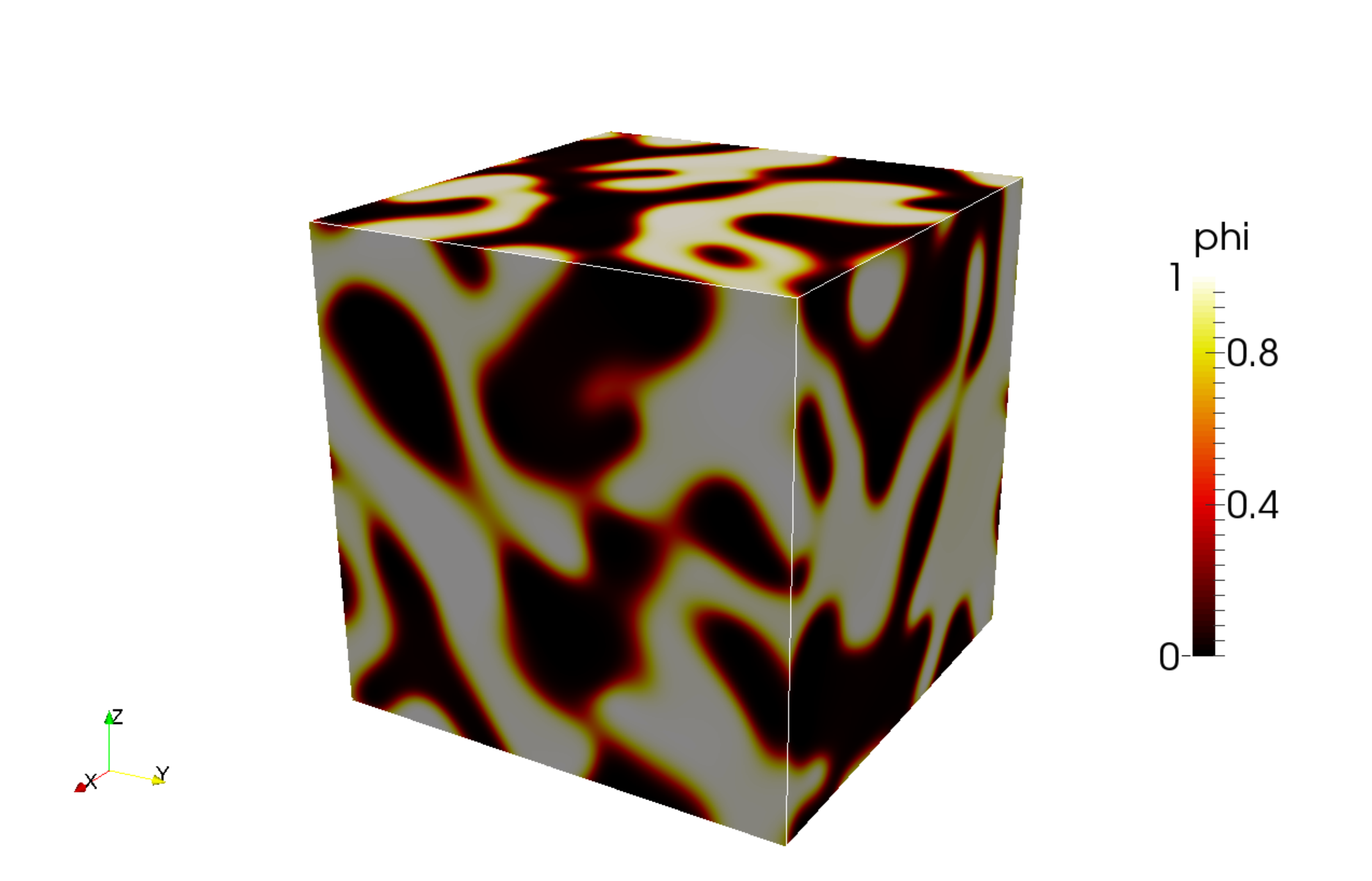}}
	\hspace{-.2in}
	\subfigure[$\,\,t=4$]{\includegraphics[width=0.32\textwidth]{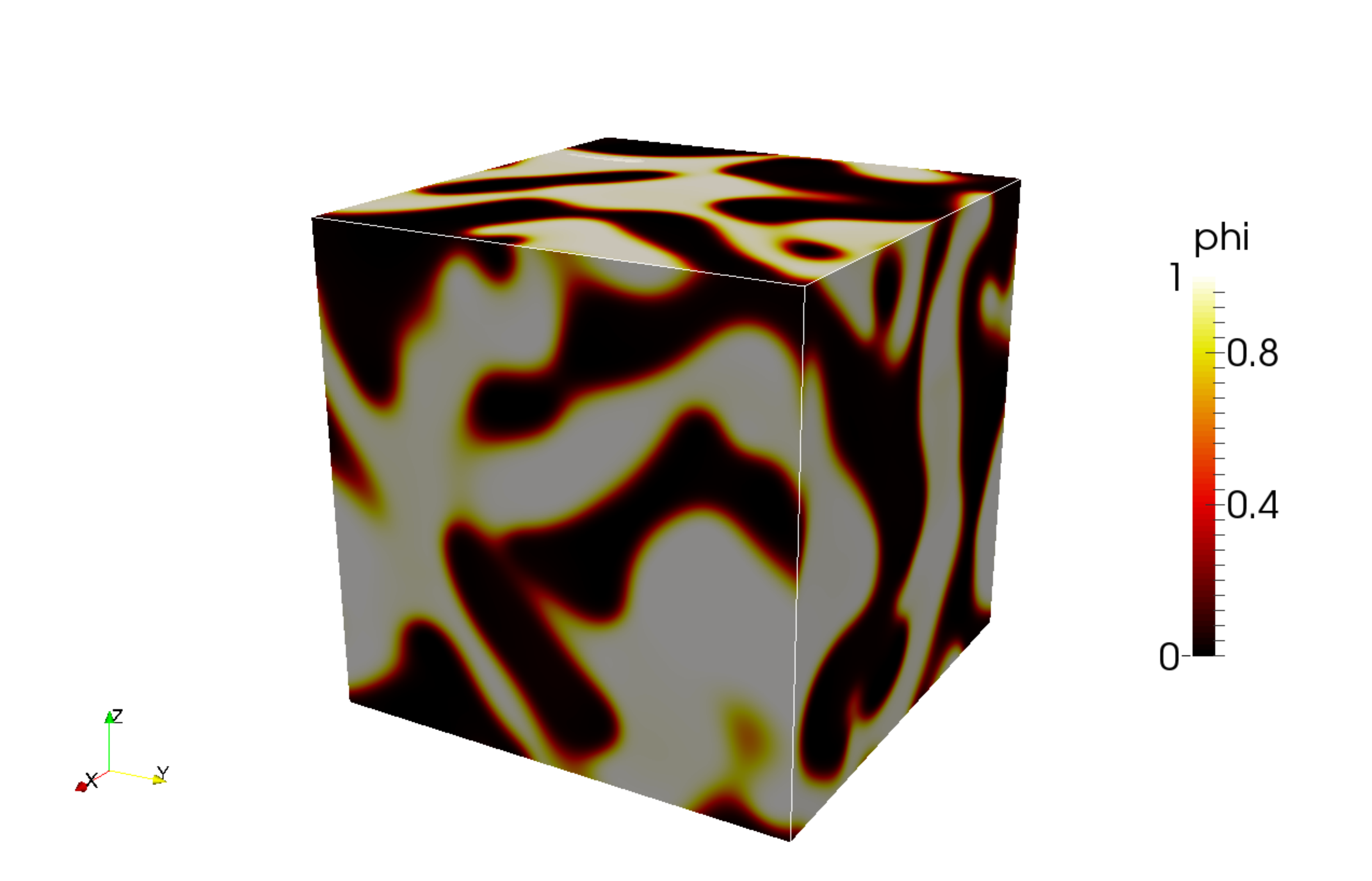}}
	\caption{Snapshot results for FDCH3D-Long.  Shown are isovolume plots based on $\phi=(C+1)/2$.  Top row: $\mydiff=10^{-5}$; middle row: $\mydiff=10^{-3}$; bottom row: $\mydiff=10^{-1}$.   }
\label{fig:snapshot_flow}
\end{figure}

Snapshots of the concentration for the case FDCH3D-Long  are shown in Fig.~\ref{fig:snapshot_flow}.  
For $\mydiff=10^{-5}$, the concentration field is well mixed and the domain structure is not discernible.  For $\mydiff=10^{-1}$ a clear domain structure emerges, albeit that the shape of the domains is modified by the flow.  The case with $\mydiff=10^{-3}$ is intermediate between these extremes.

\begin{figure}[t]
	\centering
		\subfigure[]{\includegraphics[width=0.48\textwidth]{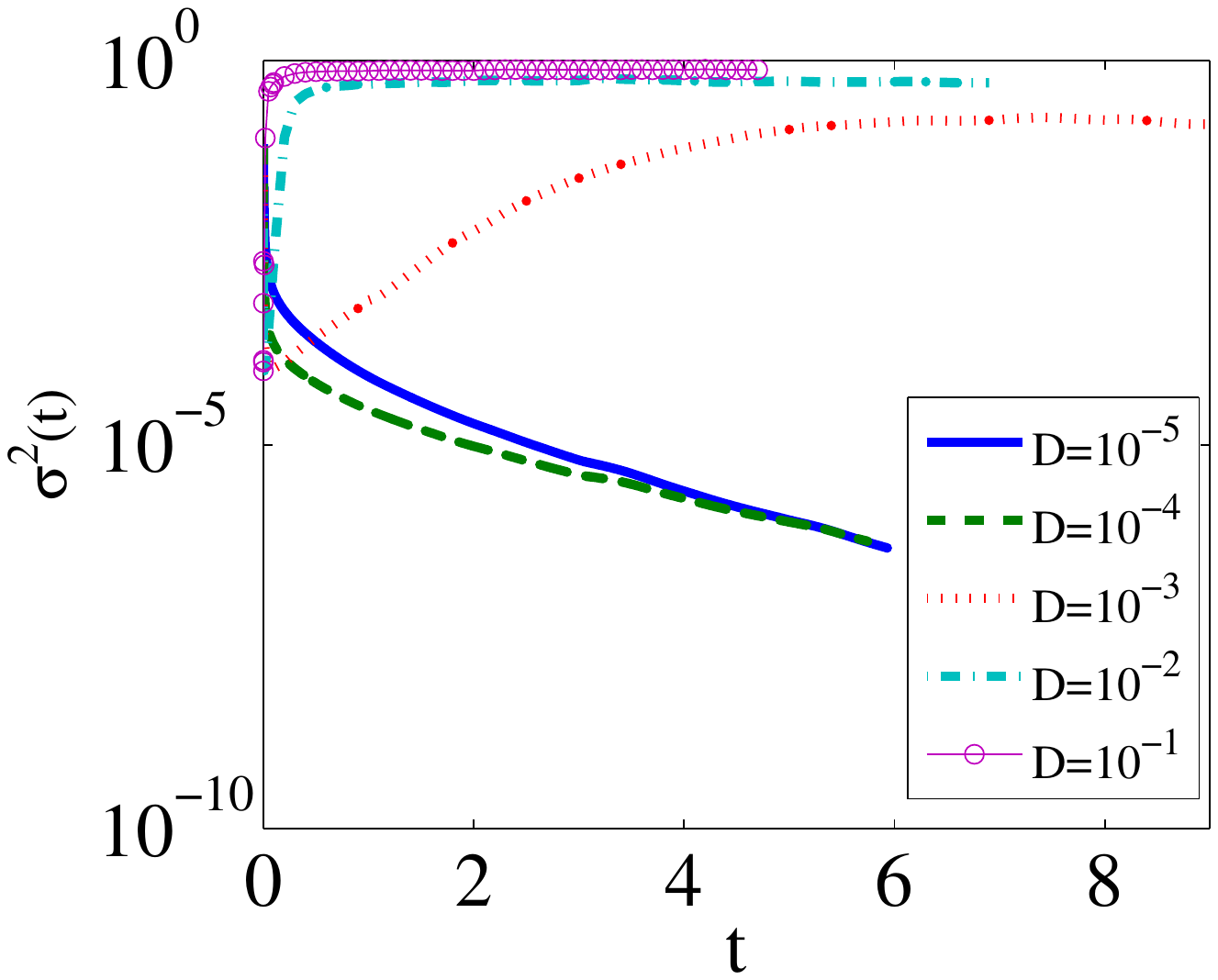}}
		\subfigure[]{\includegraphics[width=0.48\textwidth]{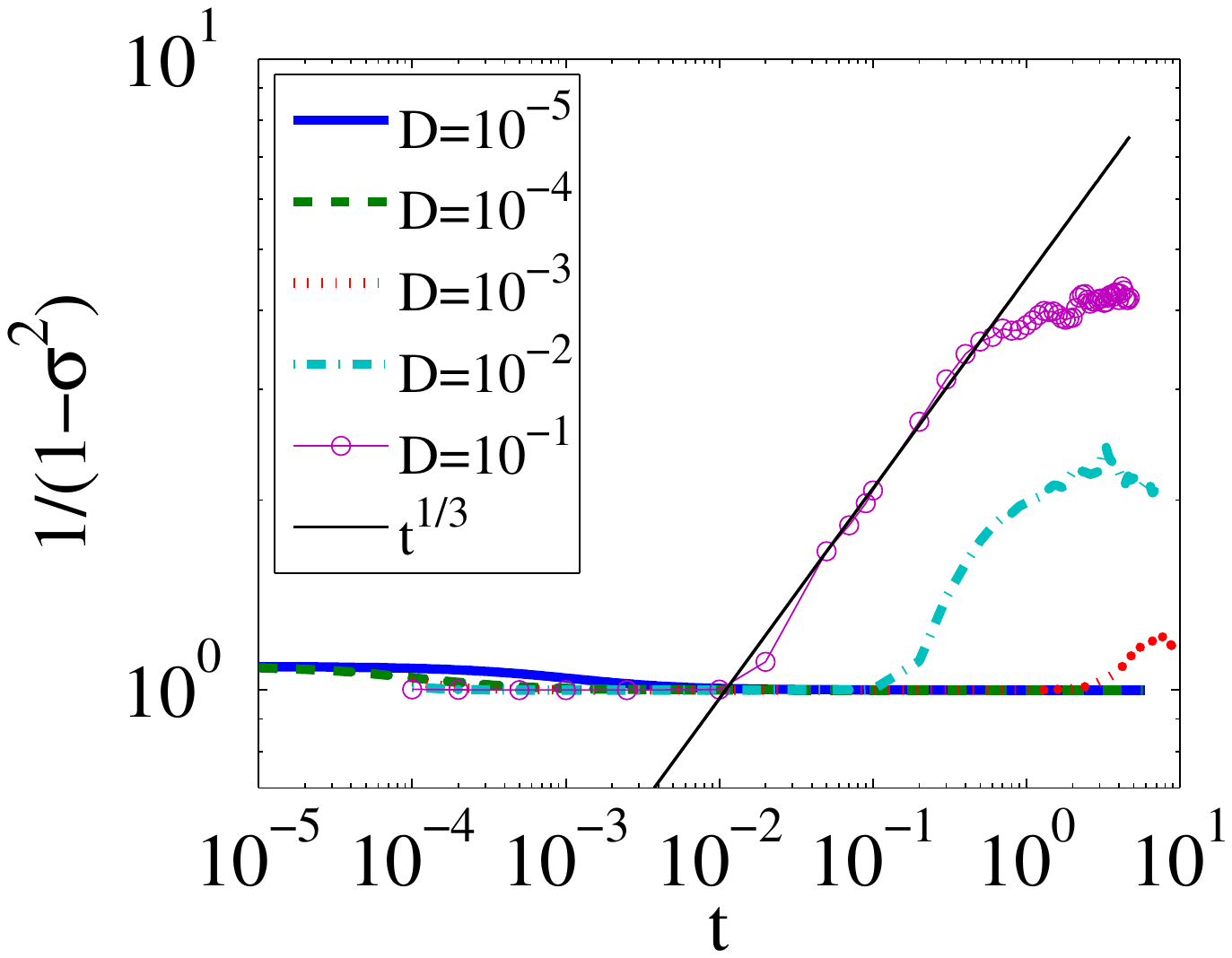}}
\caption{Results from the simulation FDCH3D-Long: (a) the concentration variance; (b) the lengthscale $\ell\propto 1/(1-\sigma^2)$, shown as a function of time for various values of the inverse P\'eclet number $\mydiff$.  }
\label{fig:time_series}
\end{figure}

The variance $\sigma^2(t)$ as a function of $\mydiff$ is examined in Fig.~\ref{fig:time_series}(a). 
For $\mydiff\leq 10^{-4}$ the variance decays exponentially in time, at a  $\mydiff$-dependent rate.    
For $\mydiff\geq 10^{-2}$ the variance grows rapidly and then saturates at a value close to unity.  The case $\mydiff=10^{-3}$ is intermediate:  the  fate of the variance is saturation, but this is a slow process, and the saturated value is much less than unity.   
Also, if the variance saturates, it can be used to measure the typical size of the binary-fluid domains, through the formula $\ell\propto (1-\sigma^2)^{-1}$ (Ref.~\cite{Berti_goodstuff} and Fig.~\ref{fig:time_series}(b)).  For $\mydiff\geq 10^{-3}$, the lengthscale grows initially as $t^{1/3}$ according to the LS growth law, until the onset of coarsening arrest induced by the shear flow. 
A similar picture emerges for the other simulations in Tab.~\ref{tab:simsx}, albeit that only in the cases FDCH3D-Long, FDCH2D-Long and 2DLA is genuine variance decay observed (in the other simulations the variance saturates at a very low level). The correlation time of the chaotic flow therefore seems to play a crucial role in the outcome of phase separation.
To summarize, three regimes can be distinguished:
\begin{enumerate}[label=(\roman*)]
\item For small values of diffusivity ($\mydiff\lesssim 10^{-4}$), the concentration variance decays exponentially, thereby leading to a well-mixed state without any discernable structure (Fig. \ref{fig:snapshot_flow}(c)). This hyperdiffusive regime has already been observed numerically in the presence of forced hydrodynamic turbulence, in the limit of high forcing amplitude \cite{Berti_goodstuff}. It was earlier proposed analytically, in the same context~\cite{Aronovitz84}, wherein it was suggested that a strong enough turbulence level might cause the system to become well-mixed through an enhancement of eddy-diffusivity;
\item For high values of diffusivity ($\mydiff\gtrsim 10^{-2}$), the characteristic lengthscale of the flow initially grows according to the LS law, then eventually saturates at a finite value, reflecting a clear domain structure (see Fig. \ref{fig:snapshot_flow}(i)). This coarsening arrest was first discussed by \cite{Ruiz81}, and first evidenced numerically in \cite{chaos_Berthier} for a chaotic flow. It was then observed and quantified in numerous works, in the presence of shear \cite{shear_Berthier,naraigh2007bubbles} or turbulent \cite{Berti_goodstuff,Perlekar2014} flows;
\item For intermediate values of diffusivity ($\mydiff\approx 10^{-3}$), the concentration variance saturates more slowly, at a value much smaller than in case (ii) (Fig. \ref{fig:snapshot_flow}(f)).
\end{enumerate}

\begin{figure}
	\centering
	\includegraphics[width=0.6\textwidth]{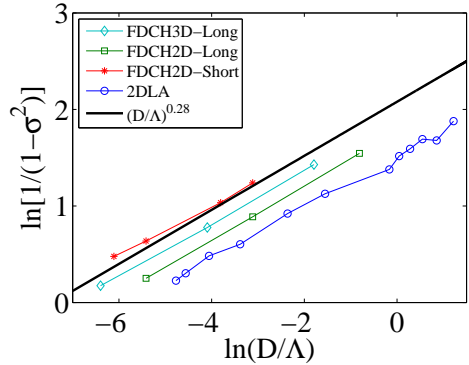}
	\caption{The time-average of $(1-\sigma^2)^{-1}$ corresponding to coarsening arrest.  }
	\label{fig:lengthscales_variance}
\end{figure}

To place these results in a more quantitative context, a comparative study based on all simulation results in Tab.~\ref{tab:simsx} is performed.  The saturated value of $(1-\sigma^2)^{-1}$ (where it exists) is plotted for the various simulations (Fig.~\ref{fig:lengthscales_variance}).  The quantity of data available for this particular plot is relatively meagre for the non-lattice-based simulations.  However, for 2DLA a large quantity of data is available.  All of the evidence suggests a dependence $\ell\propto (D/\Lambda)^{0.28}$ for the arrested lengthscale, both in two and three dimensions.   
%
%

The coarsening arrest can be explained as a dynamic balance between the advection term $\bm{u}\cdot\nabla C$ with typical magnitude $\mylep$ and the phase-separation term $\mydiff\nabla^2(C^3-C-\gammanondim\nabla^2 C)$, with typical magnitude $\gammanondim^{1/2}\mydiff/\ell^3$ (Ref.~\cite{Bray_advphys}), leading to a balance $\ell \sim (\mydiff/\mylep)^{1/3}$~\cite{Berti_goodstuff,shear_Berthier}.   The measured exponent (common to all simulations in both two and three dimensions) is $0.28$ and is  therefore close to the theoretical value.  In Ref.~\cite{Berti_goodstuff} scaling exponents between $0.28$ and $0.29$ are reported for flow fields completely distinct from the configuration in Eq.~(\ref{eq:flow_def},\ref{eq:flow_def_2d}), thereby highlighting the applicability of the `universal' theory for which the exponent is precisely $1/3$.

\subsection{Onset of the diffusive regime: interpretation in terms of Batchelor lengthscales} \label{subsec:res_onset_diff}
\label{subsec:diff_scales}

\begin{figure}
	\centering
	\subfigure[]{\includegraphics[width=0.45\textwidth]{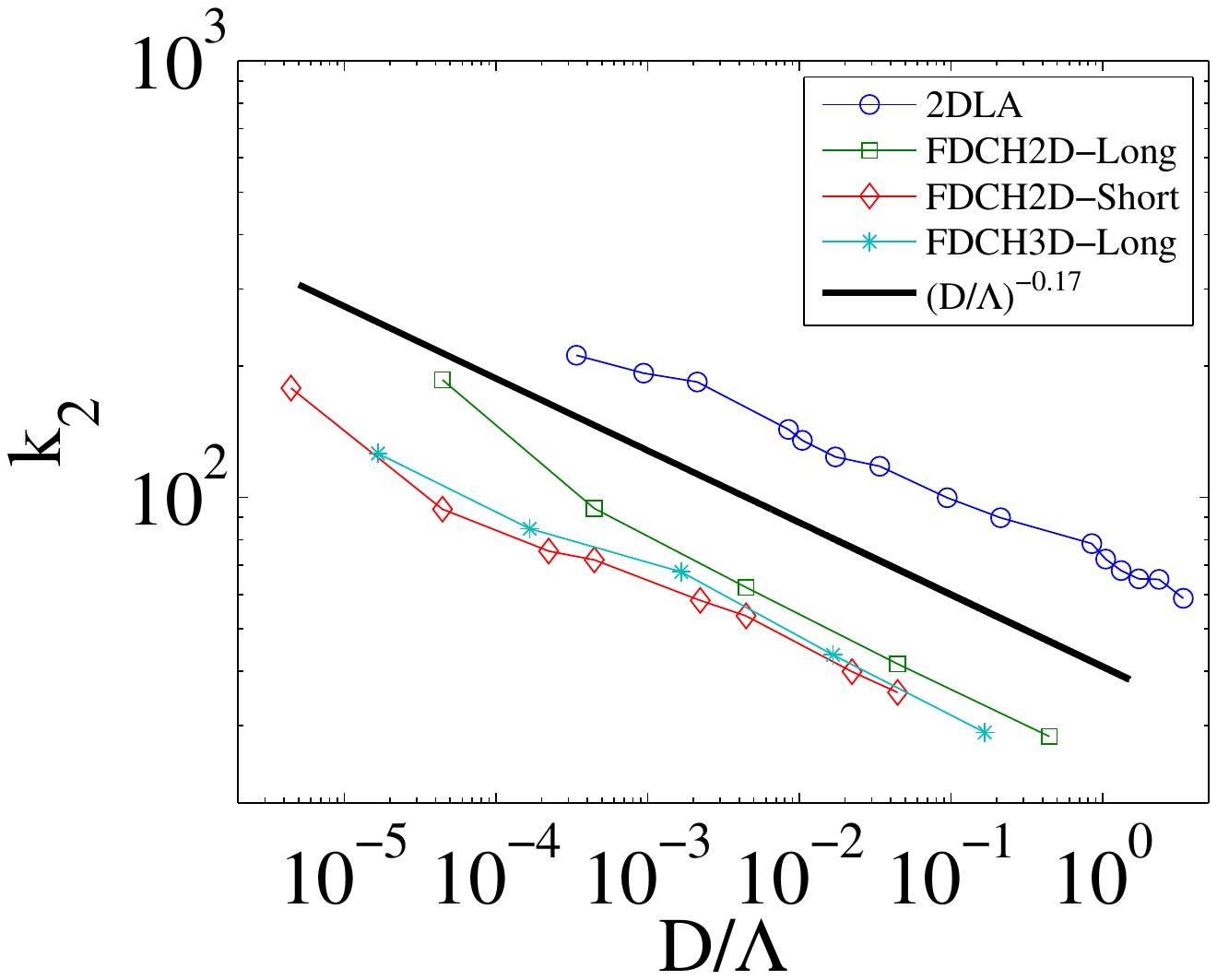}}
	\subfigure[]{\includegraphics[width=0.45\textwidth]{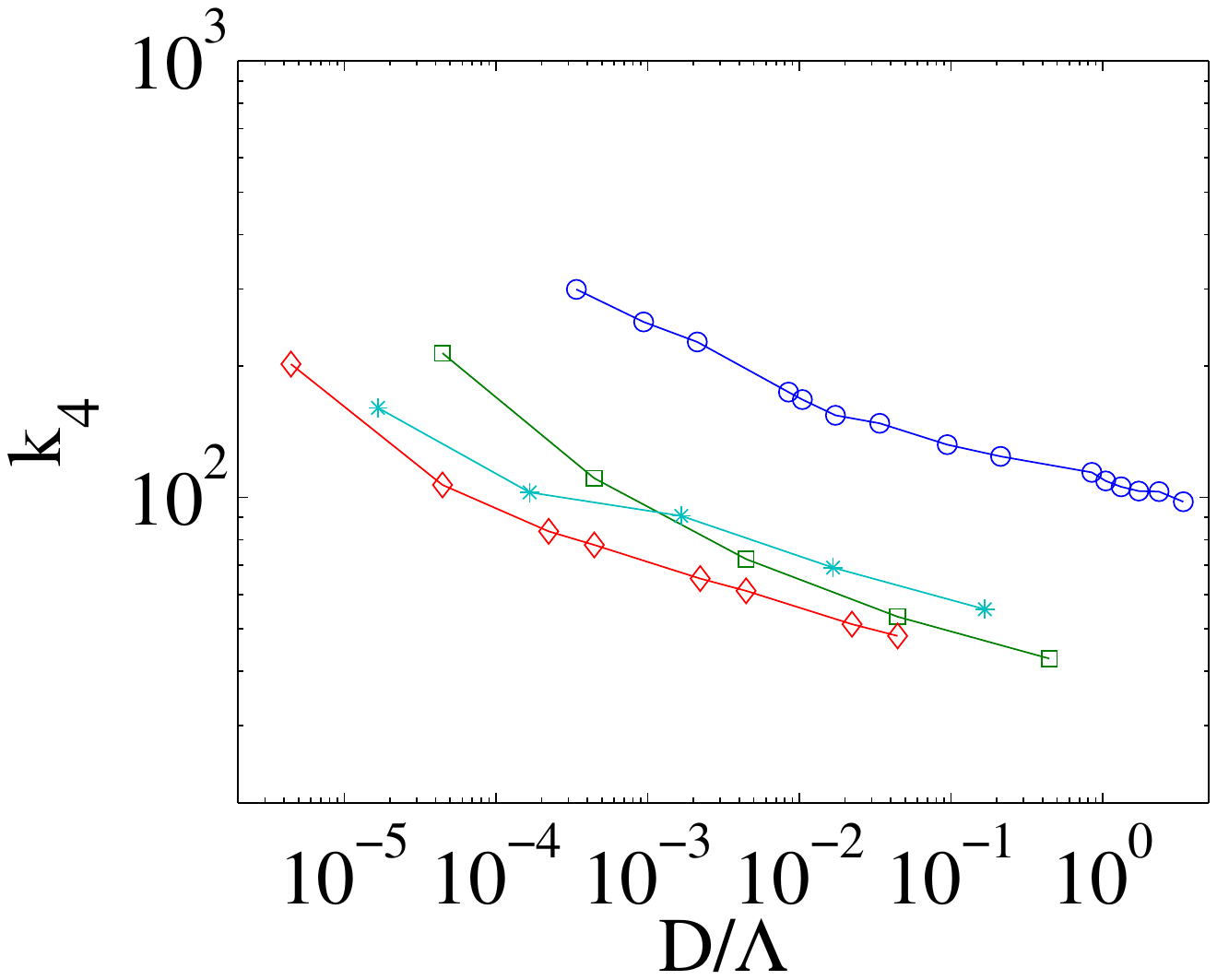}}
	\caption{The time-average of the Batchelor scales $k_2$ (a) and $k_4$ (b).  Plot key: (circles) 2DLA; (squares) FDCH2D-Long; (diamonds) FDCH2D-Short; (stars) FDCH3D-Long.  The case FDCH3D-Short is not shown because convergence to a statistically-steady state was not reached by $t=10$. }
	\label{fig:all_scales}
\end{figure}

The chosen measure of lengthscale loses its relevance when the variance decays, coinciding with a regime wherein hyperdiffusion is important in its own right, thereby leading to a collection of disparate scales that enter the balance between advection and the other terms in Eq.~\eqref{eq:ch_flow}.
For this reason, consideration is given also to the Batchelor lengthscales $k_2=(\overline{\|\nabla C\|_2^2/\|C\|_2^2})^{1/2}$ and $k_4=(\overline{\|\nabla^2 C\|_2^2/\|C\|_2^2})^{1/4}$
%
%
time averaged over (quasi-steady) late times.  The results are shown in Fig.~\ref{fig:all_scales}.  For the Batchelor scale $k_2$, it is possible to fit the curve $k_2\sim (\mydiff/\mylep)^{-0.17}$ through all the data in Tab.~\ref{tab:simsx}, where the exponent is obtained by taking the average over all the simulation families (average: ${-0.17}$, standard deviation $0.02$).  Similarly, for the scale $k_4$ a clear trend is visible among all the simulation families wherein the dependence on $(\mydiff/\mylep)$ changes as one moves from coarsening arrest to a more diffusive regime accompanied by a decay in the concentration variance.  The trend is clearest in the data-rich 2DLA simulation runs and for that reason, a power-law fit for $k_4$ is presented in that context only, with $k_4\sim (\mydiff/\mylep)^{-0.09}$ for coarsening arrest and $k_4\sim (\mydiff/\mylep)^{-0.16}$ for diffusion.  Fits for the other simulation cases yielded a similar result albeit that some variation exists between each simulation family.  This is not surprising: there is a larger quantity of data in the 2DLA simulations and the rapidity of the simulations performed resulted in averages being taken over long  intervals.  Nevertheless, a clear similarity between all the simulation runs is evidenced by Fig.~\ref{fig:all_scales}.   Again focusing on 2DLA, a further simulation run was performed wherein it was found that $k_{2,4}\sim \gammanondim^{-0.34}$.

These scaling results can be used to obtain a more   quantitative criterion for the onset of variance decay.  A variance budget is derived
by  multiplying Eq.~\eqref{eq:ch_flow} by $C$ and averaging the result over the spatial domain $[0,L]^n$:
\begin{equation}
\tfrac{1}{2}\frac{d\myvar^2}{dt}=-\mydiff\langle (3C^2-1)|\nabla C|^2\rangle-{\gammanondim \mydiff\langle |\nabla^2C|^2\rangle},
\label{eq:var_budget}
\end{equation}
where the second term of the right-hand side is negative definite and represents dissipation; the first term is sign-indefinite and can correspond either to variance production or dissipation.  The focus herein is on the decaying case:  assuming a decaying concentration variance, quartic terms on the right-hand side of Eq.~\eqref{eq:var_budget} are neglected compared to quadratic terms, yielding $d\sigma^2/dt\approx 2\mydiff\langle |\nabla C|^2\rangle-2\gammanondim \mydiff\langle |\nabla^2C|^2\rangle$, which
%
%
is further approximated as
\begin{equation}
\frac{d\myvar^2}{dt}\approx 2\mydiff\left(k_2^2-\gammanondim k_4^4\right)\myvar^2,
\label{eq:var_budget2}
\end{equation}
and the variance growth rate is estimated as $\rho=2\mydiff\left(k_2^2-\gammanondim k_4^4\right)$, with $\myvar^2\propto \mathe^{\rho t}$.
The remaining terms in Eq.~\eqref{eq:var_budget2} are estimated using the scaling laws for the Batchelor scales:
%
%
$k_2= c_2 (\mylep/{\mydiff})^p\gammanondim^{-r}$ and 
$k_4= c_4 (\mylep/{\mydiff})^q\gammanondim^{-r}$,
where $c_2$ and $c_4$ are $O(1)$ positive constants independent of the physical parameters.  The onset of variance decay is therefore given by $\rho=0$, hence $k_2^2=\gammanondim k_4^4$.  Using the scaling rules for $k_2$ and $k_4$,
this condition can be re-expressed as $\mylep/\mydiff=(c_4^4/c_2^2)^{1/(2p-4q)}\gammanondim^{(1-2r)/(2p-4q)}$.
%
%
A momentary reversion to dimensional physical parameters suggests that the natural balance for the onset of hyperdiffusion is $\mylep/\mydiff\propto \gammanondim^{-1}$, further suggesting $2r-1=2p-4q$.  For the simulation database given by 2DLA, we have $2p-4q=2(0.17)-4(0.16)=-0.30$ and  $2r-1=-0.32$.  Allowing for some spread between these measured exponents and the true values,  the theory can be regarded as self-consistent.
The theory is also consistent with the scaling argument given in Ref.~\cite{naraigh2007bubbles} for the concentration spatial structure to contain diffusive filaments rather than the droplets more characteristic of phase separation, albeit that the decay of the concentration variance was not observed in that earlier work owing to the authors' not having probed the full range of flow amplitudes and diffusivities.

\subsection{Probability distribution function of the concentration in the hyperdiffusive regime} \label{subsec:res_PDF}
\label{subsec:pdf}

The difference between the two- and three-dimensional cases is clearest in the hyperdiffusive regime. 
For this reason, we examine the probability distribution function (PDF) of the concentration, taken at late times, as a function of the parameter $\mydiff$.  Consideration is given to the cases FDCH3D-Long, FDCH2D-Long and 2DLA wherein genuine variance decay observed.  Each simulation family produces qualitatively similar results: the PDF is unimodal and centred at the origin in the hyperdiffusive regime, while for the coarsening-arrest regime the distribution is bimodal, with phase separation $C=\pm 1$ being favoured (see, \textit{e.g.}, Ref.~\cite{naraigh2007bubbles}).  Between these extremes there exists a crossover distribution that is unimodal and centred at the origin but having a width that is much larger than the earlier purely diffusive cases.

\begin{figure}
	\centering
	\subfigure[]{\includegraphics[width=0.32\textwidth]{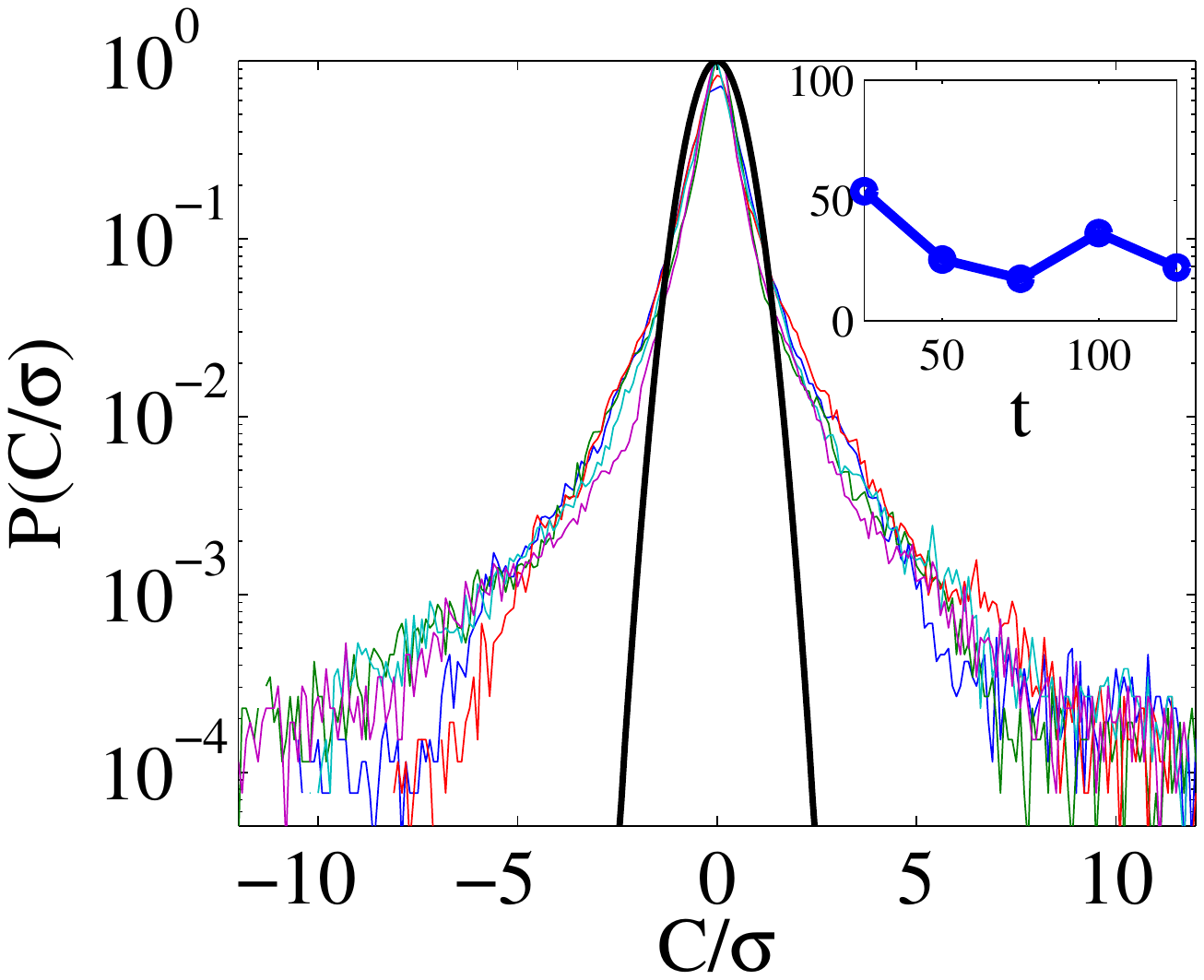}}
	\subfigure[]{\includegraphics[width=0.32\textwidth]{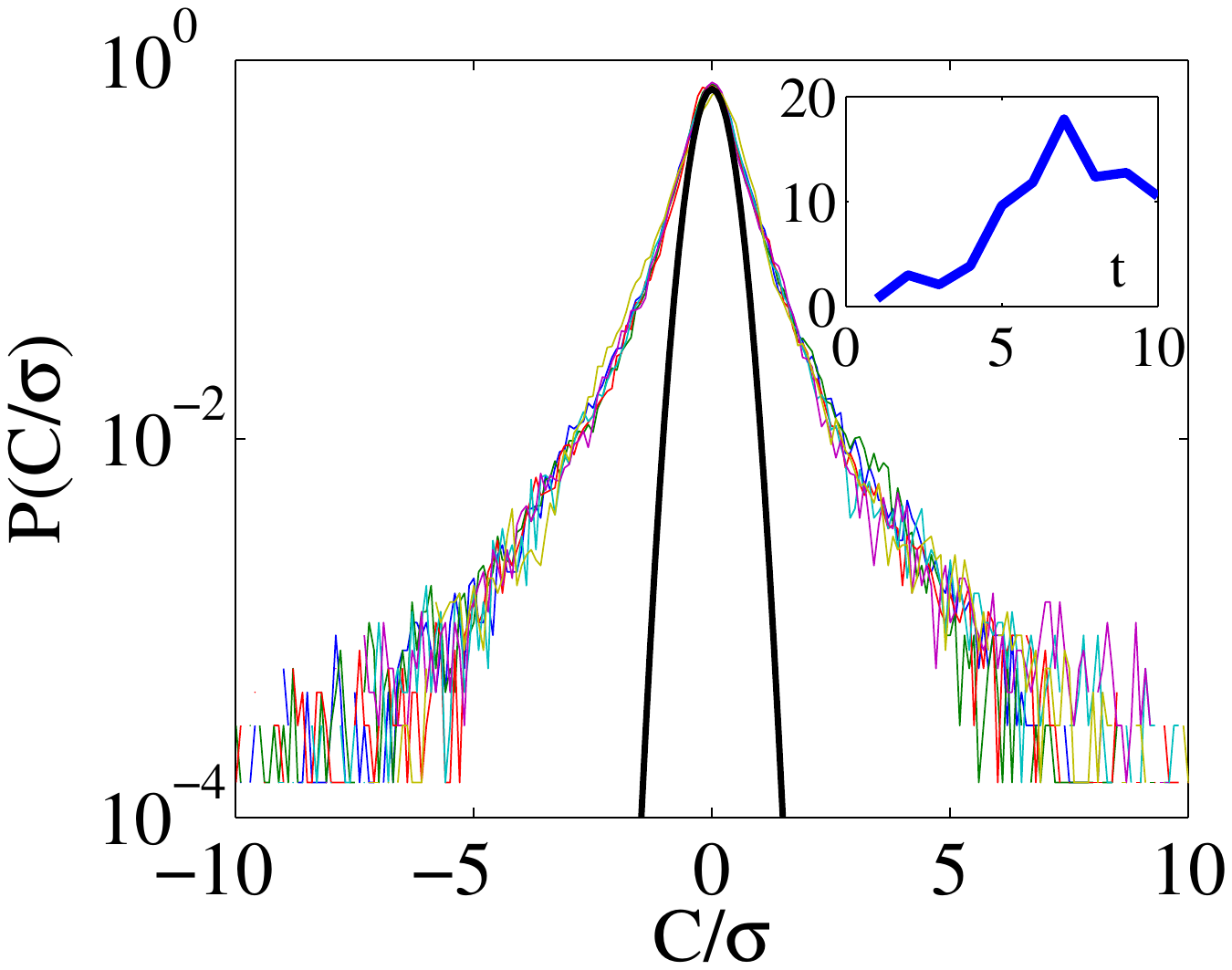}}
	\subfigure[]{\includegraphics[width=0.32\textwidth]{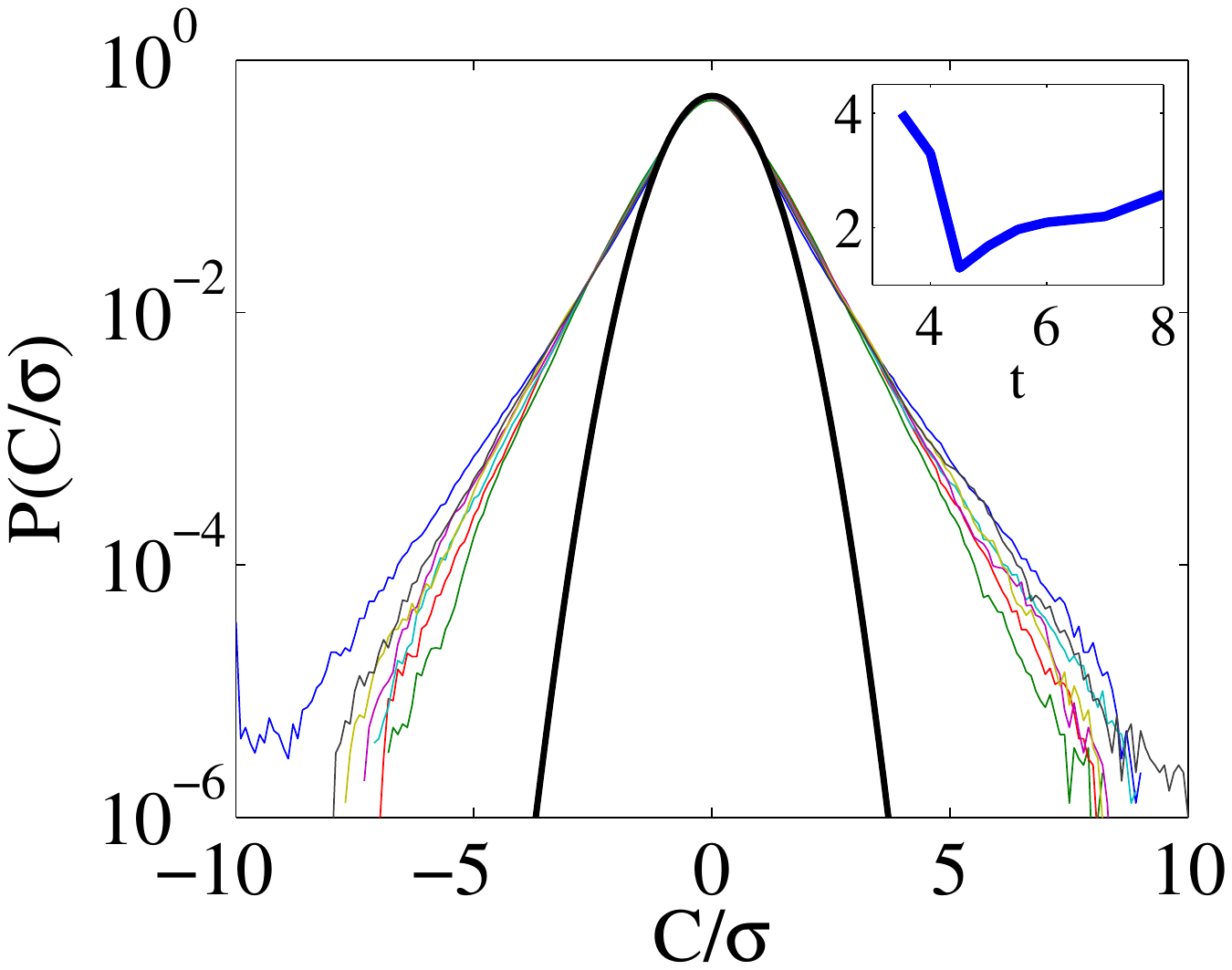}}
	\caption{PDF of concentration for decaying cases: (a)  a particular long-correlation-time realisation of 2DLA, with $\tau_{\mathrm{corr}}=0.5$ and $\mydiff=2\times 10^{-3}$,  (b) FDCH2D-Long, $\mydiff=10^{-5}$, $t\in[4,10]$; (c) FDCH3D-Long, $\mydiff=10^{-5}$, $t\in[4,8]$.  The inset in each panel shows the kurtosis defect (`flatness') of the PDF as a function of time. }
	\label{fig:hist_selfsimilar}
\end{figure}

Further examination of the PDF for hyperdiffusive regimes and at late times is presented in Fig.~\ref{fig:hist_selfsimilar}.  The variable along the horizontal axis is $C/\myvar(t)$.
The two-dimensional studies attain a self-similar distribution during the course of the simulations, such that the kurtosis defect of the distribution is constant on average (Fig. \ref{fig:hist_selfsimilar}(a) and (b)).  The PDF possesses a Gaussian core, followed by an exponentially-decaying region just outside the core.  The tails of the PDF decay more slowly however, corresponding to a very flat distribution and the large kurtosis defect seen in the inset panels in Fig.~\ref{fig:hist_selfsimilar}(a) and (b).  In other words, the normalized PDF possesses `fat tails' whereby extreme events are relatively common (compared to random extreme events associated with Gaussian processes).  Thus, pockets of unmixed binary fluid where the variance is still quite high are likely.

The simulation FDCH3D-Long also demonstrates a statistically steady Gaussian core followed by an exponentially-decaying region just outside the core (Fig. \ref{fig:hist_selfsimilar}(c)).  The kurtosis defect increases over the course of the simulation, meaning that further flattening of the distribution is expected as time goes by, indicating also that a completely statistically-steady state is not attained.  Nevertheless, the core and exponentially-decaying regions of the distribution exhibit self-similarity out to five standard deviations.  From these results it is possible to draw a contrast between the two- and three-dimensional cases: in the two-dimensional case, a fully self-similar PDF is attained relatively rapidly, characterized by an extremely fat-tailed distribution.  The tendency towards equilibrium is slower in three dimensions and the distribution (while still far from normal) has narrower tails.  

Mathematically, this result makes sense, at least under the supposition  that the effects of the advection can be parametrized at very late times by a linear theory involving an effective-diffusion operator.
Certainly, in the regime of interest the amplitude of the concentration is maintained at small values and linearized dynamics pertain.  Although effective-diffusion theory may not apply exactly (even accounting for the presence of the Cahn--Hilliard hyperdiffusion term), such an approach gives an insight into the effect of dimensionality.  Thus, we use here a diffusive-type process to model the late-stage linearized dynamics.
The correlation function $\mathrm{Corr}(\vecr,t)=L^{-d}\int_{[0,L]^d}\mathd^d x\, C(\vecx+\vecr)C(\vecx)$ for such linearized dynamics on a periodic domain is readily calculated, and such calculations (along standard lines) reveal that $\mathrm{Corr}(\vecr,t)$ decays more rapidly in three dimensions rather than two as $|\vecr|\rightarrow\infty$: correlations persist over longer scales in 2D compared to 3D, leading to fatter tails in the concentration PDF in 2D compared to 3D~\cite{brockwell2009time}.  More precise work along these lines will be the subject of future work (\textit{cf.} References~\cite{Haynes2005,Balkovsky1999} for pertinent calculations going beyond effective-diffusion theories but relating only to pure advection-diffusion processes).
Indeed, the stationarity of the normalized PDF is reminiscent of the `strange eigenmode' for advection-diffusion~\cite{lattice_PH2}, wherein the spatial distribution of the tracer concentration in a chaotic flow (with correlation times comparable to the shear timescale) becomes statistically self-similar, modulo the constraint that the variance decays exponentially.
In contrast to advection-diffusion however, the variance budget and the associated heuristic arguments advanced herein indicate a non-trivial dependence of the variance decay rate on the diffusivity.  These distinctions nevertheless open up the possibility that a non-trivial extension of the spectral theory of advection-diffusion~\cite{Haynes2005} might apply to the present (linearized) large-flow-amplitude advective Cahn--Hilliard dynamics. 

\section{Flow anisotropy in the coarsening arrest regime} 
\label{sec:res_aniso}

\begin{figure}[htb]
	\centering
  \includegraphics[width=0.45\textwidth]{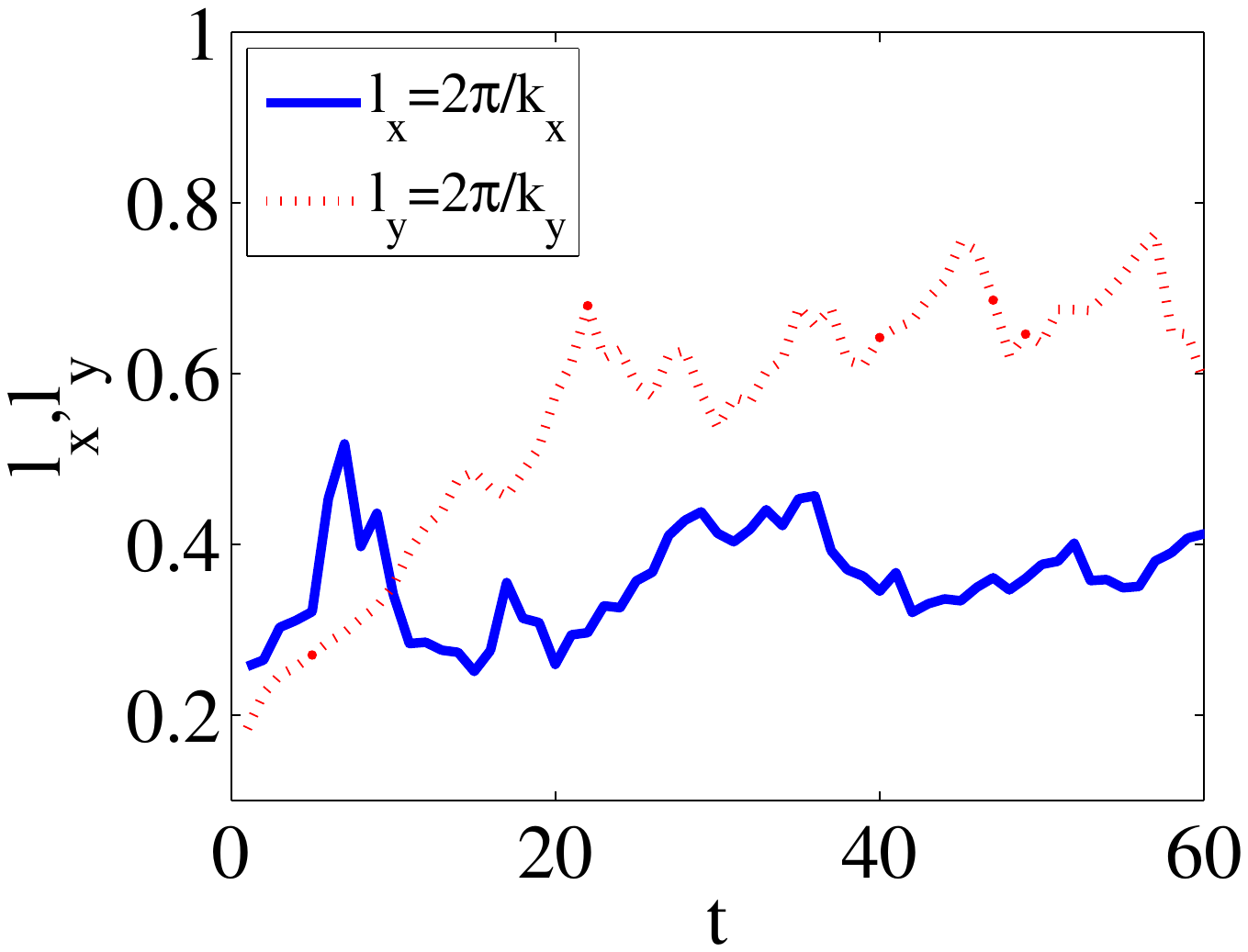}
\caption{Results for the simulation 2DAniso2 $(N=1)$.}
	\label{fig:x2DAniso2}
\end{figure}

\begin{figure}[htb]
	\centering
	\hspace{-.2in}
	\subfigure[$\,\,t=5$]{\includegraphics[width=0.32\textwidth]{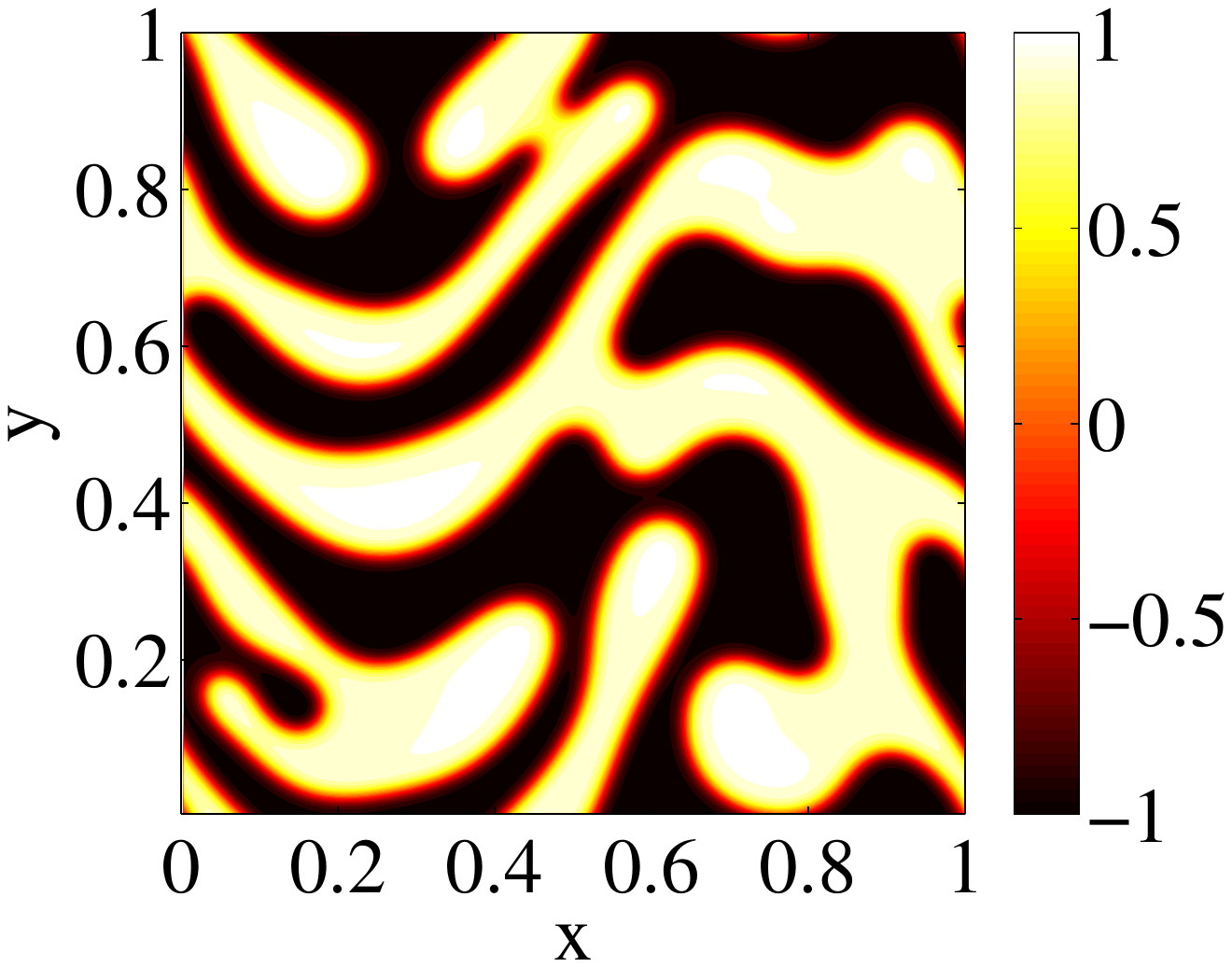}}
	\hspace{-.2in}
	\subfigure[$\,\, t=10$]{\includegraphics[width=0.32\textwidth]{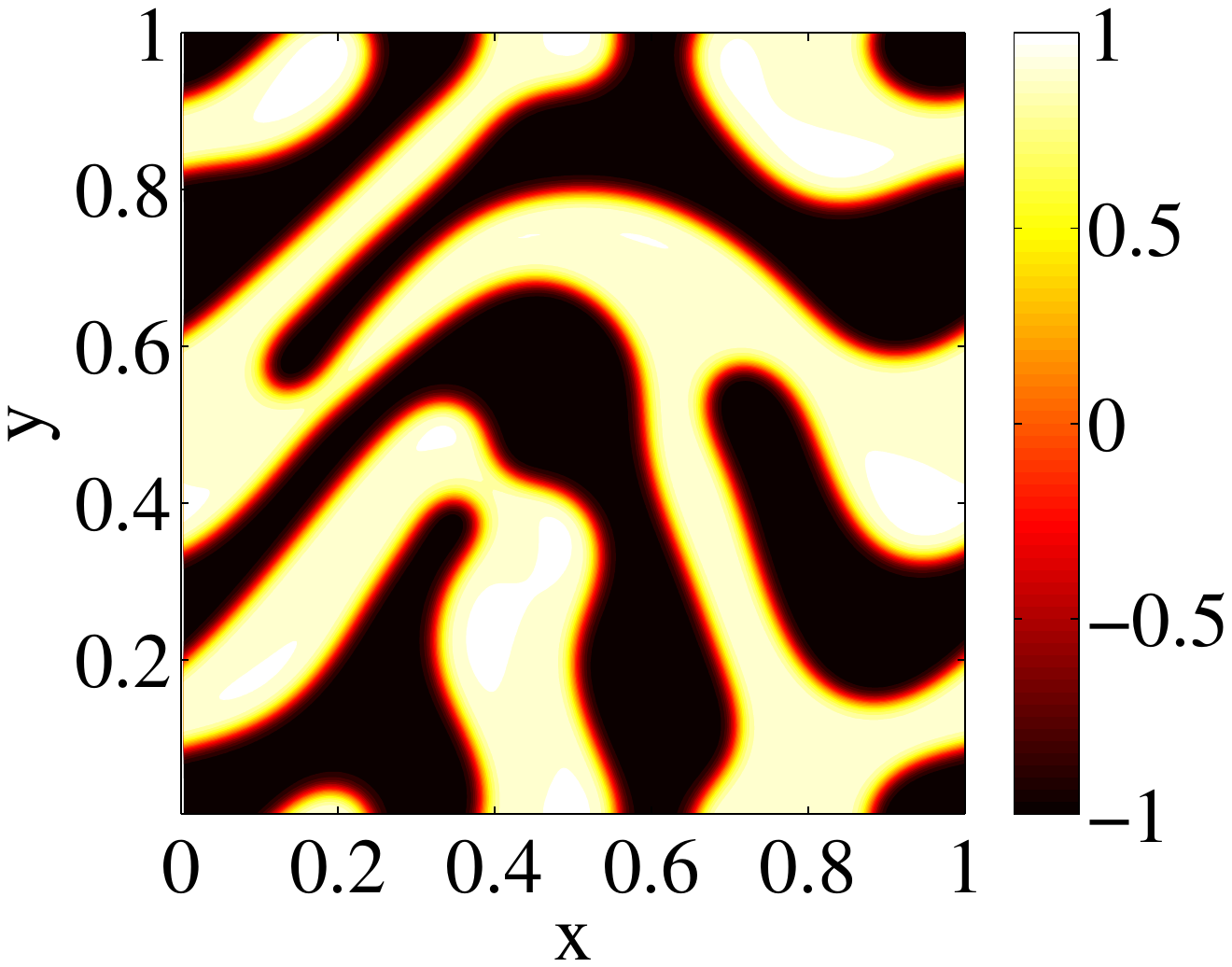}}
	\subfigure[$\,\, t=15$]{\includegraphics[width=0.32\textwidth]{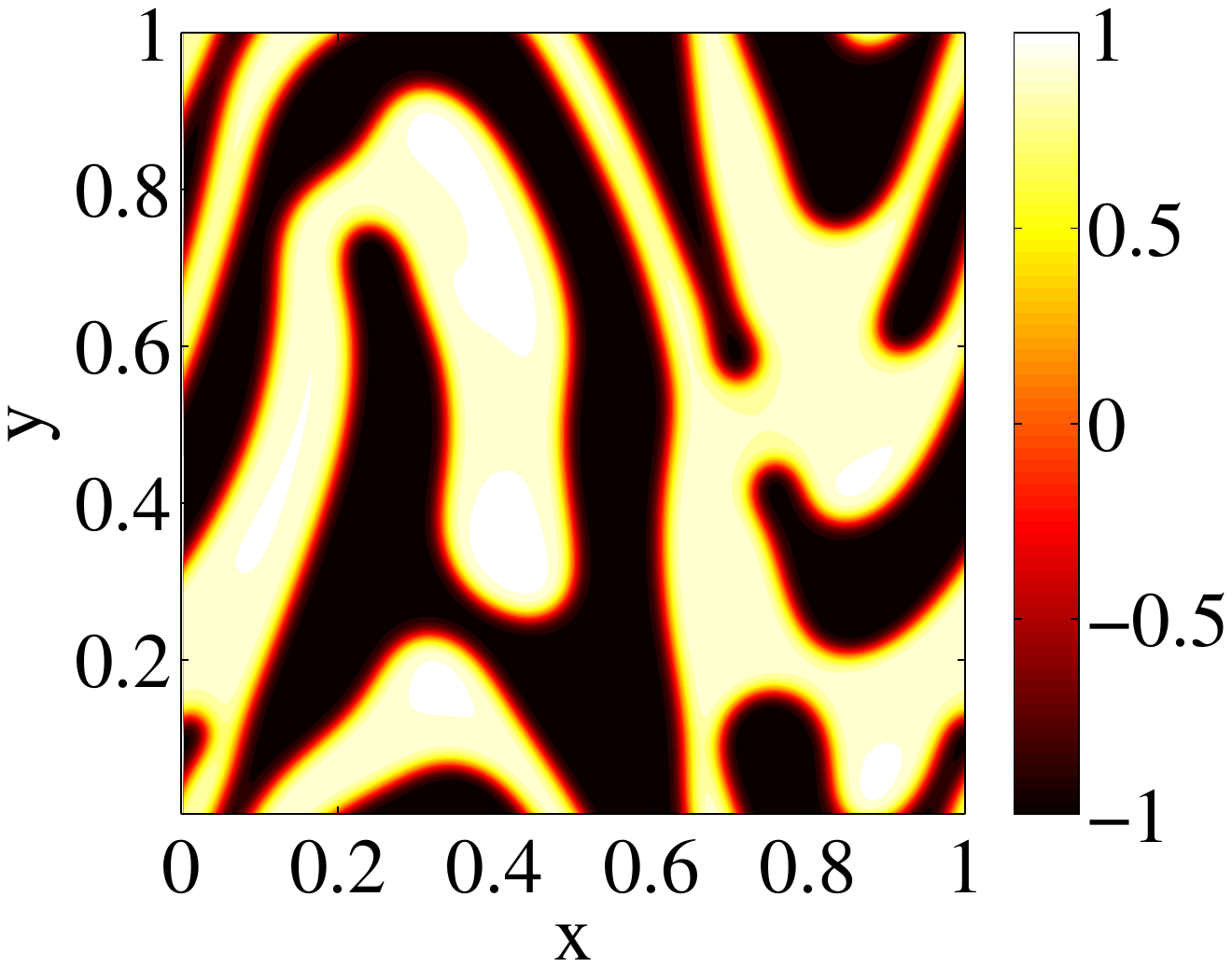}}
	\hspace{-.2in}
	\subfigure[$\,\,t=20$]{\includegraphics[width=0.32\textwidth]{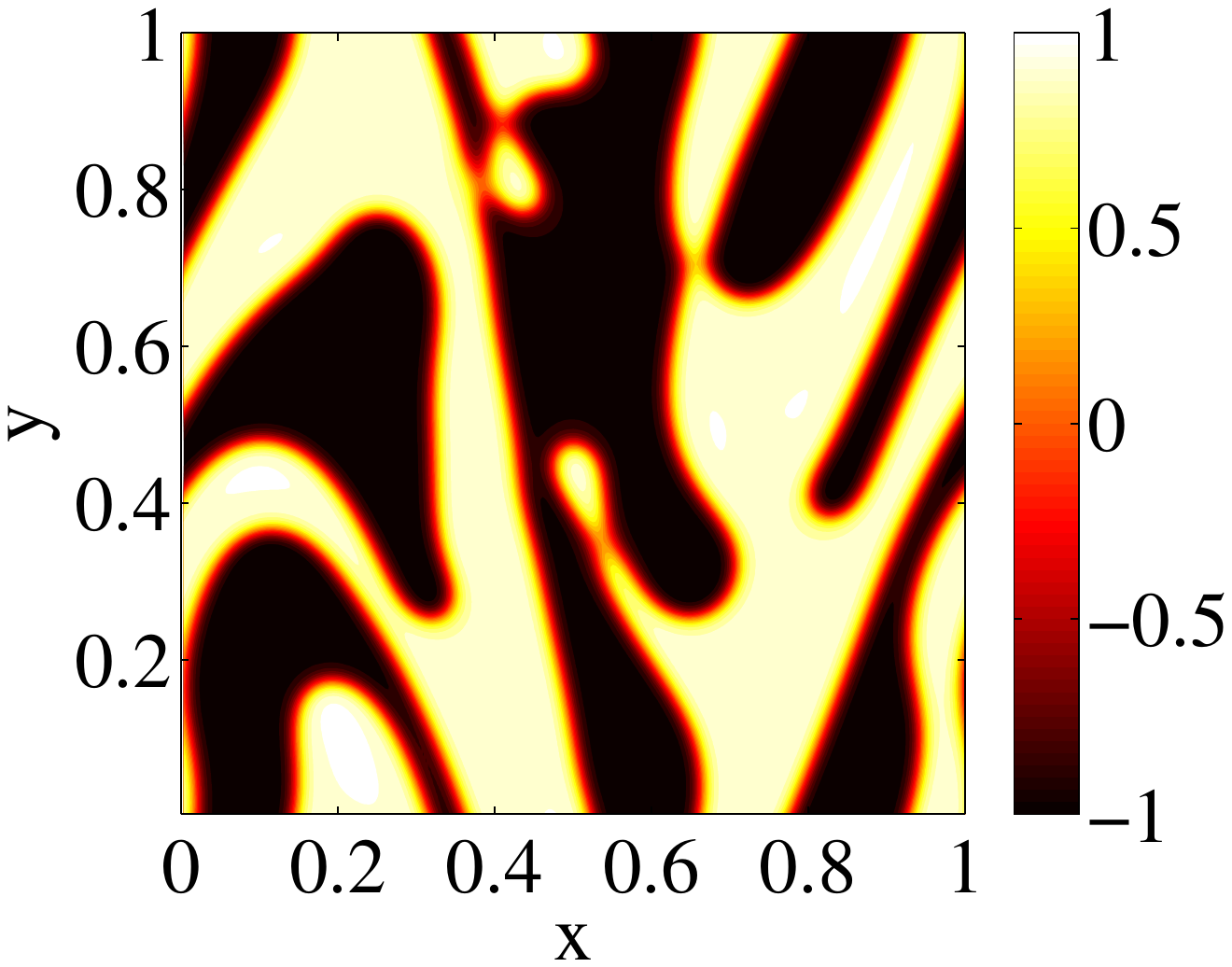}}
	\hspace{-.2in}
	\subfigure[$\,\,t=25$]{\includegraphics[width=0.32\textwidth]{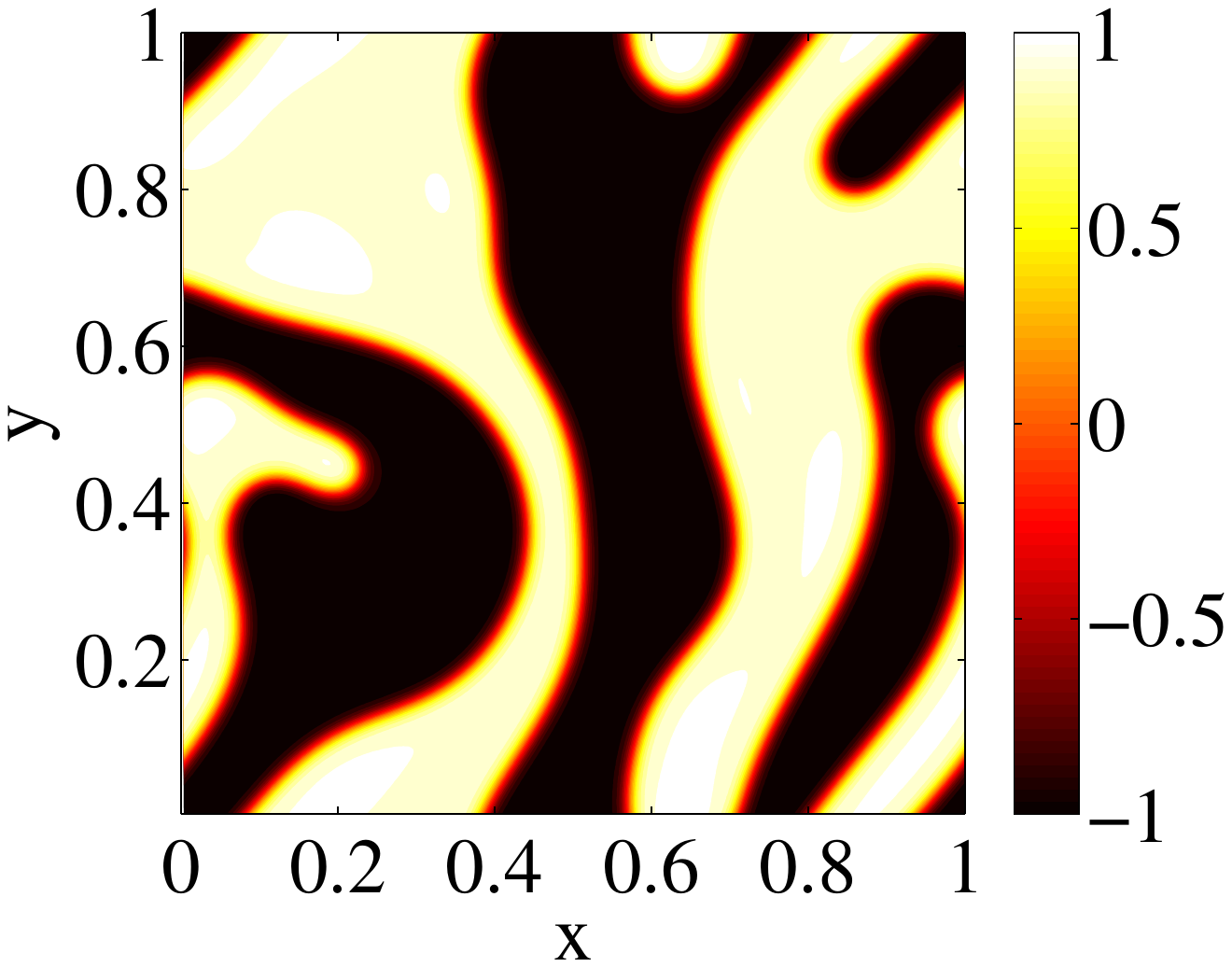}}
		\caption{Results for the simulation 2DAniso2 $(N=1)$. }
	\label{fig:2DAniso2}
\end{figure}

\begin{table}[h]
\begin{center}
		\begin{tabular}{|c|c|c|c|c|}
		\hline
			     &    Grid Size & $N$ & $\tau_\mathrm{corr}=N\tau$ & $\Delta t$\\
			\hline
		  \hline	
			2DAniso1,$\,\,$3DAniso1   & $314^n$ & 100& 1 & $10^{-4}$\\
			2DAniso2,$\,\,$3DAniso2   & $314^n$ & 1  &  0.01 &$10^{-4}$\\
			\hline
		\end{tabular}
\caption{Simulations with anisotropic flow and the code FDCH, with $\mydiff=10^{-2}$ and $\gammanondim=10^{-4}$.  For the 3D cases, $A_x=A_y=\sqrt{3/2}\sin(\pi/12)$ and $A_z^2=3-A_x^2-A_y^2$ corresponding to a larger flow amplitude in the $z$-direction.  For the 2D cases, $A_x=\sqrt{2}\sin(\pi/12)$ and $A_y^2=2-A_x^2$, corresponding to a larger flow amplitude in the $y$-direction.}
\label{tab:sims_aniso}
\end{center}
\end{table}
The random-phase sine flow~\eqref{eq:flow_def} is reparametrized such that the flow amplitudes in the three spatial dimensions are possibly distinct and equal to $A_x$, $A_y$, and $A_z$ (an analogous procedure is carried out with respect to the two-dimensional random-phase sine flow~\eqref{eq:flow_def_2d}).  The aim again is to compare and contrast the phase separation in two and three dimensions.  Four simulations are performed, the details of which are summarized in Tab.~\ref{tab:sims_aniso}.  The results are analysed in what follows.
The 2D results are discussed first.  
\begin{figure}[htb]
	\centering
		\subfigure[]{\includegraphics[width=0.45\textwidth]{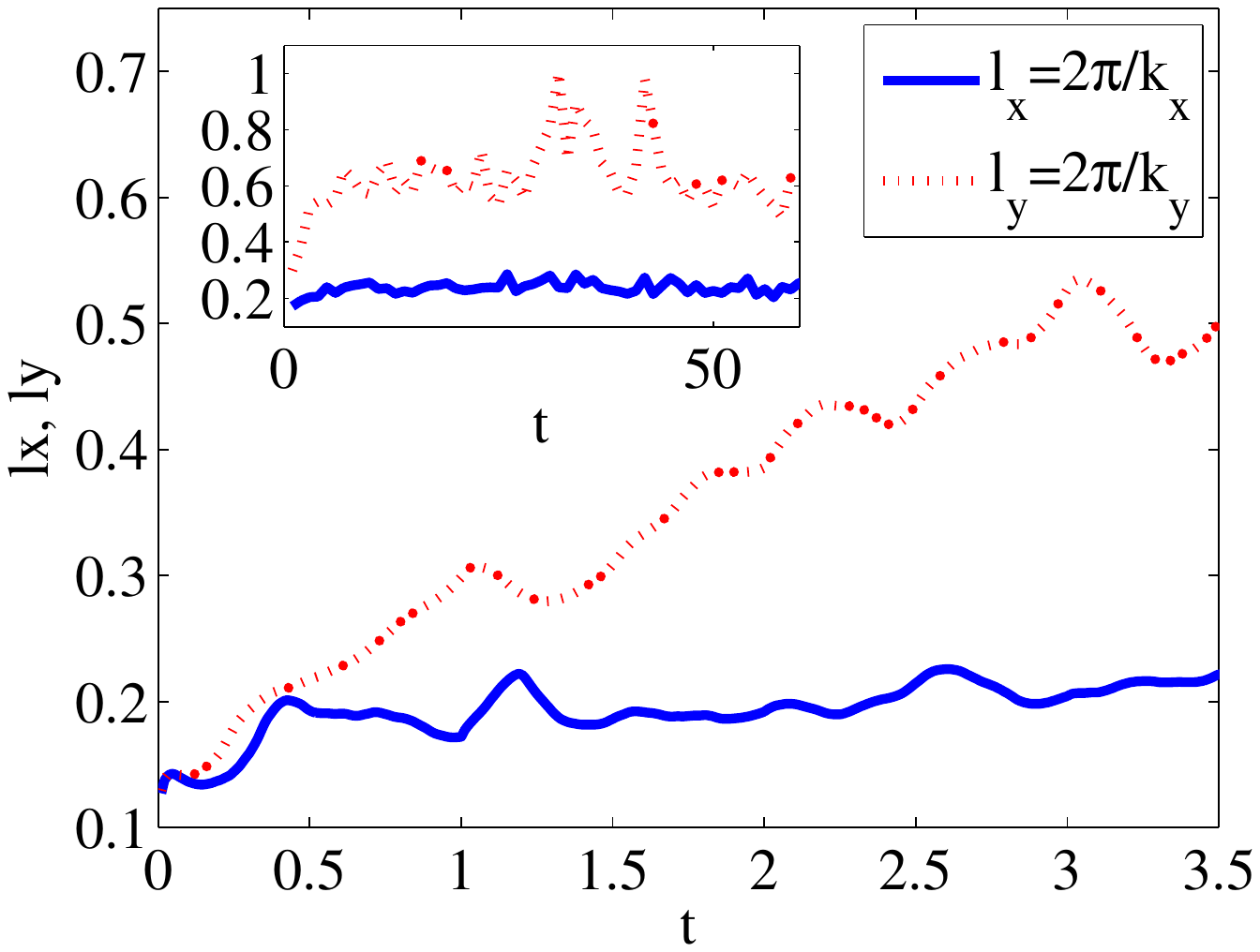}}\\
	\subfigure[$\,\,t=0.5$]{\includegraphics[width=0.34\textwidth]{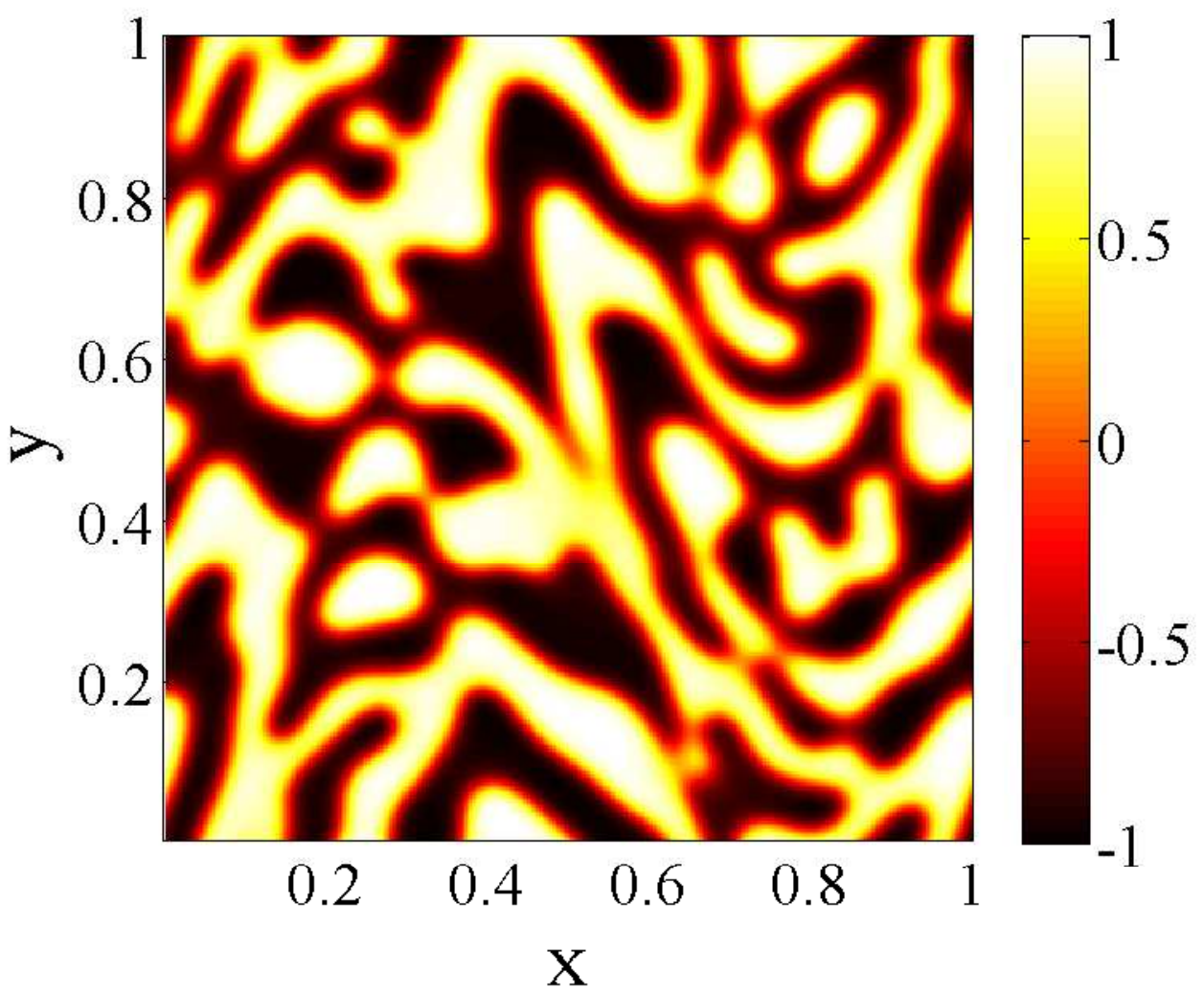}}
	\subfigure[$\,\,t=5$]{\includegraphics[width=0.34\textwidth]{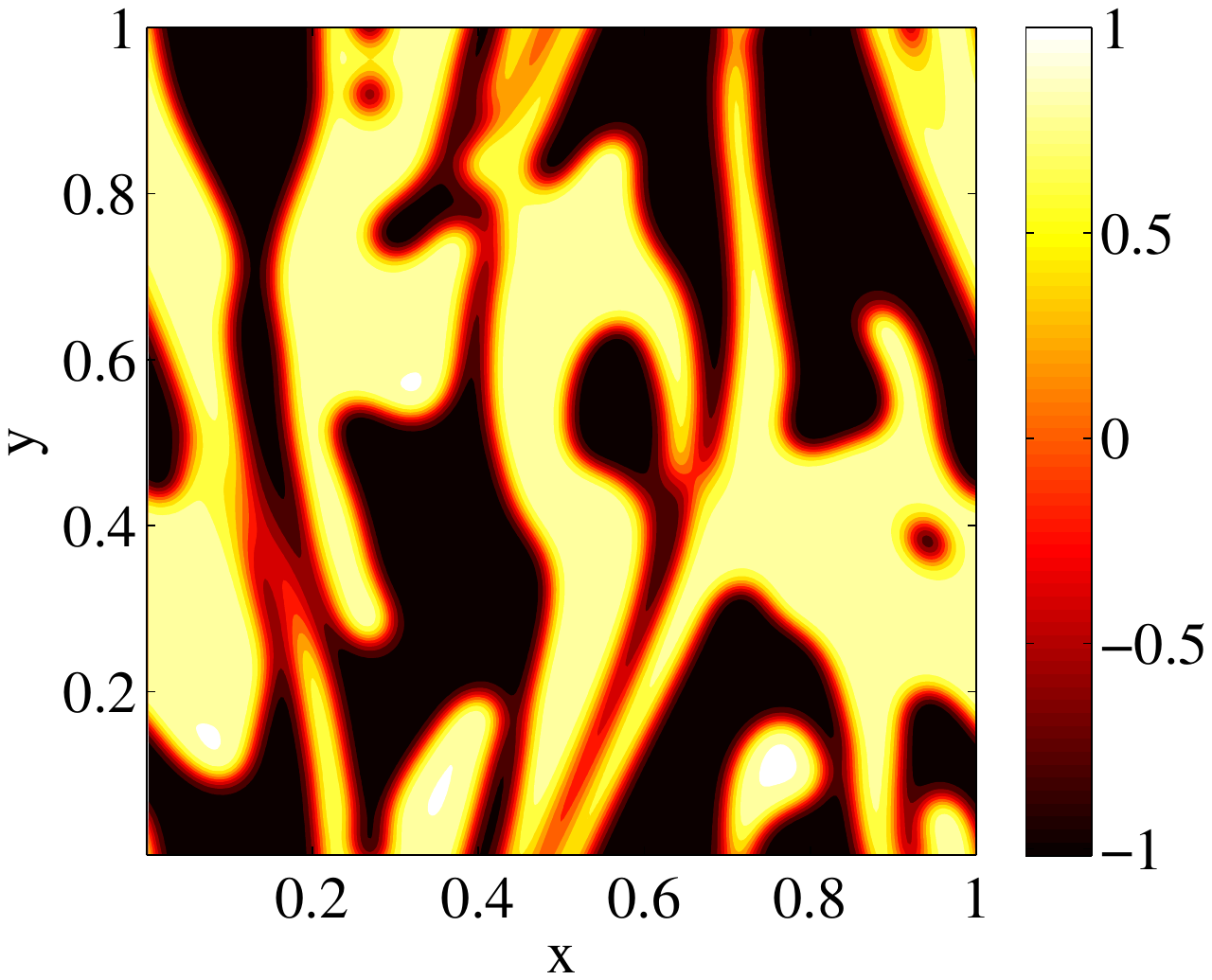}}
		\caption{Results for the simulation 2DAniso1 $(N=100)$.  (a) Batchelor lengthscales as a function of time (the inset is the same plot, only shown over a much longer time interval). (b,c) Snapshots of concentration: (b) just before saturation of $\ell_x$, (c) after saturation of $\ell_x$.}
	\label{fig:2DAniso1}
\end{figure}
The Batchelor scales 
\begin{equation}
k_{x,y}(t)=\frac{\sum_{\veck} k_{x,y}|C_{\veck}(t)|^2}{\sum_{\veck}|C_{\veck}(t)|^2},\qquad \ell_x=2\pi/k_x,\qquad \ell_y=2\pi/k_y
\end{equation}
are obtained as a measure of the typical lengthscales in each spatial direction, and indicate that for long correlation times, the domains align rapidly in the direction of \textit{greatest} flow amplitude (Fig.~\ref{fig:2DAniso1}).  In contrast, for short correlation times, the domains align transiently in the direction of \textit{least} flow amplitude before slowly re-aligning in the opposite direction such that the asymptotic state ends up the same regardless of correlation time (Fig.~\ref{fig:x2DAniso2} and \ref{fig:2DAniso2}).
Further simulations on larger domains (not shown) reveal that the saturation is not due to finite-size effects, although such effects do alter the time of saturation.

These results are explained as follows.  In the case of long correlation times, the transport is advective.  Power counting on the equation of motion gives $1/T=A_x/\ell_x$ in the first half-period of the random-phase sine flow and $1/T=A_y/\ell_y$ in the second half-period of the flow.  The timescale is assumed to be the same in each half-period: the domains adjust to the flow such that the time-scale of the Cahn--Hilliard dynamics can be equated with $T$ in the above equations, giving $A_x/\ell_x=A_y/\ell_y$, hence $\ell_y/\ell_x=A_y/A_x$, hence $\ell_y>\ell_x$.  This prediction agrees qualitatively with Figure~\ref{fig:2DAniso1} and Figures~\ref{fig:x2DAniso2}, \ref{fig:2DAniso2} (the latter at late time), although the precise quantitative agreement is absent.  This is not surprising, as the estimates for the timescale in the above equations are only heuristic: better estimates might be obtained by the kind of Lagrangian calculations performed at the beginning of Section~\ref{sec:res} for the computation of the (isotropic) Lyapunov exponent $\Lambda$.

In the case of short correlation times, the constant renovation of the phases disrupts the coherence of the advective transport, meaning the effect of the flow is more diffusive in nature~\cite{pavliotis2008multiscale}.  Additionally, at short times, the amplitude of the concentration field is small, meaning that a linearized dynamics is appropriate, wherein the advective term is parametrized by an effective diffusivity, with
$\partial_t\llangle C\rrangle=\partial_i\left(\mathcal{D}_{\mathrm{eff},ij}\partial_j \llangle C\rrangle\right)-\mydiff\nabla^2\llangle C\rrangle-\mydiff\gammanondim\nabla^4\llangle C\rrangle$, where $\llangle C\rrangle
$
denotes the homogenized concentration field in the regime of the linearized dynamics, and where $\mathcal{D}_{\mathrm{eff},ij}$ is the effective diffusivity tensor. Due to the choice $A_y>A_x$ in the flow field, in a first approximation we take $\mathcal{D}_{\mathrm{eff},ij}=\mathcal{D}_\mathrm{eff}\delta_{i2}\delta_{j2}$, to give
\begin{equation}
\frac{\partial}{\partial t}\llangle C\rrangle=\mathcal{D}_\mathrm{eff}\frac{\partial^2}{\partial y^2}\llangle C\rrangle
-\mydiff\nabla^2\llangle C\rrangle-\mathcal{C}_\mathrm{n}\mydiff\nabla^4\llangle C\rrangle,
\end{equation}
for which the dispersion relation is
\begin{equation}
\sigma(k_x,k_y)=\mydiff(k_x^2+k_y^2)-\gammanondim\mydiff(k_x^2+k_y^2)^2-\mathcal{D}_\mathrm{eff}k_y^2,
\label{eq:sig_eff}
\end{equation}
with a maximum along the $x$-axis.  Thus, the modes selected by the transient linearized dynamics are oriented strictly along the $x$-direction. This description agrees qualitatively with the picture of the early-time dynamics in Figure~\ref{fig:2DAniso2}: here, the early-time dynamics of the domain-formation are `frozen' in to the concentration field at early times up to $t=5$ (such that the domains align in the $x$-direction), until the nonlinear dynamics take over and lead to the eventual alignment of the domains in the $y$-direction. 

The validity of this description has been confirmed by a number of numerical experiments: we first of all carried out a numerical simulation similar to 2DAniso2, but with the nonlinear term $C^3$ in the chemical potential set to zero (i.e. linearized dynamics for all time).  In this scenario, the domains (to the extent that such structures exist within the linearized dynamics) align in the $x$-direction for all time.  A second experiment was performed identical to 2DAniso2 but with highly nonlinear initial conditions (a disc-shaped domain of $C=1$ in a sea of $C=-1$).  In this scenario, the domain is stretched in the $y$-direction from the very beginning of the simulation: no alignment (no matter how brief) in the $x$-direction is seen, confirming the importance of the linearized dynamics in the creation of transient structures aligned in the $x$-direction.

These results are consistent with our knowledge of phase separation in passive shear flows~\cite{shear_Bray,berthier2001_unishear}.  There, consideration is given to unidirectional shear flow in two dimensions (Reference~\cite{shear_Bray} also contains some discussion about the three-dimensional case), with $\vecv(\vecx)=Sy\hat{\bm{x}}$, where $S$ is the shear rate. Simple power counting on the equation of motion gives $\ell_x=St\ell_y$ in two dimensions, meaning that the system coarsens more rapidly in the flow direction.  Clearly, the present application does not correspond exactly to unidirectional shear flow but the analogy persists: the domains in the present application align in the direction of greatest flow amplitude, as opposed to the direction of the velocity gradient.
Of concern in the work on unidirectional shear flows in two dimensions is the ultimate fate of a binary fluid under passive unidirectional shear flow~\cite{shear_Bray,berthier2001_unishear}. These works reveal that $\ell_x\sim t$  and $\ell_y\sim \mathrm{Const.}$ as $t\rightarrow\infty$ (i.e. no quasi-steady state exists).  Obviously, this particular aspect is distinct from the present work, wherein the chaoticity of the flow arrests the coarsening (the coarsening arrest in Figures~\ref{fig:2DAniso1} and~\ref{fig:2DAniso2} is confirmed to be independent of finite-size effects, in the sense that  such arrest also occurs on domains of much larger size for which $\lim_{t\rightarrow\infty}(\ell_x,\ell_y)\ll L$, where $L$ denotes the box size in the simulation).

\begin{figure}[t]
	\centering
		\subfigure[$\,\,N=100$]{\includegraphics[width=0.45\textwidth]{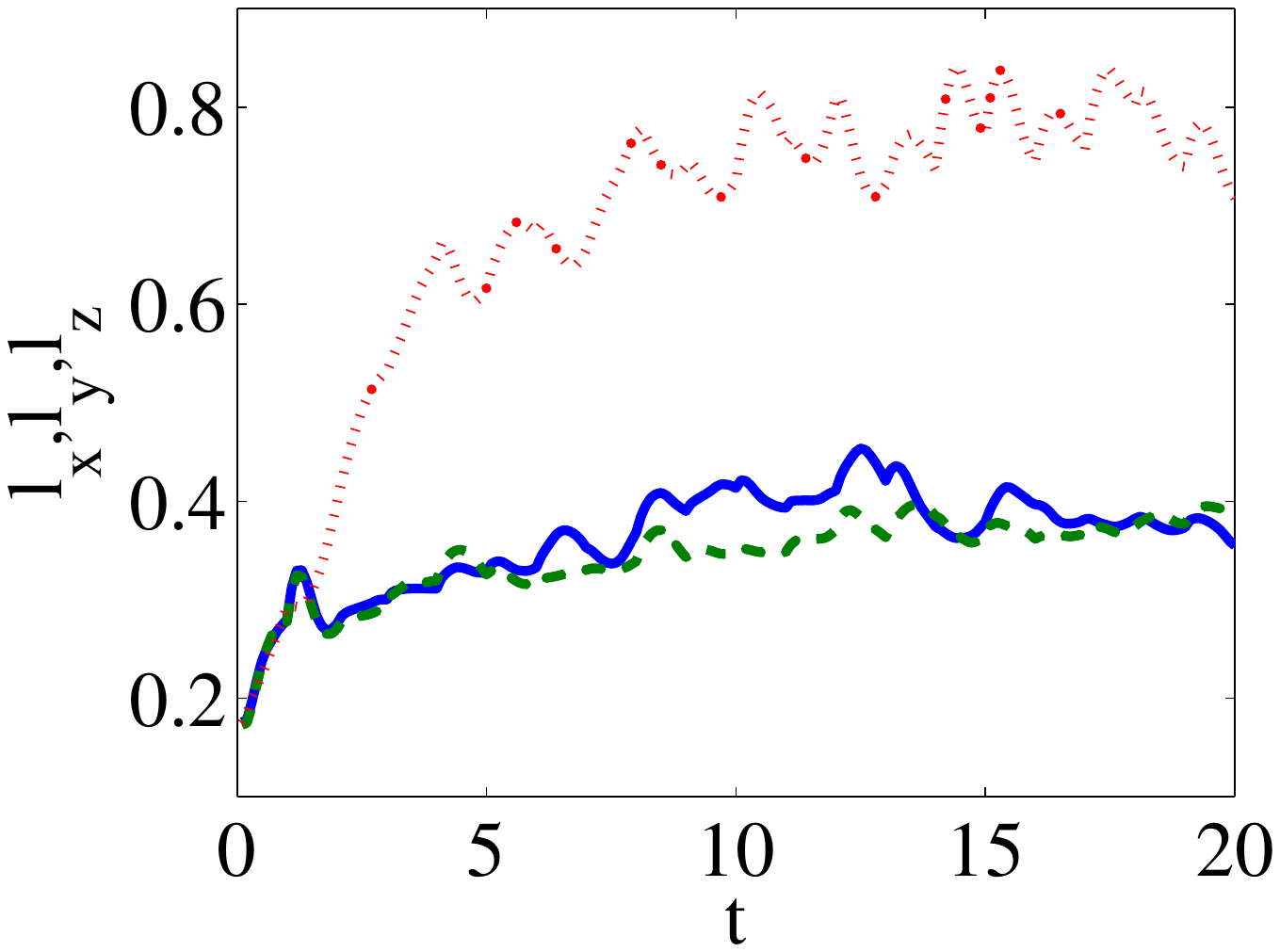}}
	  \subfigure[$\,\,N=1$]{\includegraphics[width=0.45\textwidth]{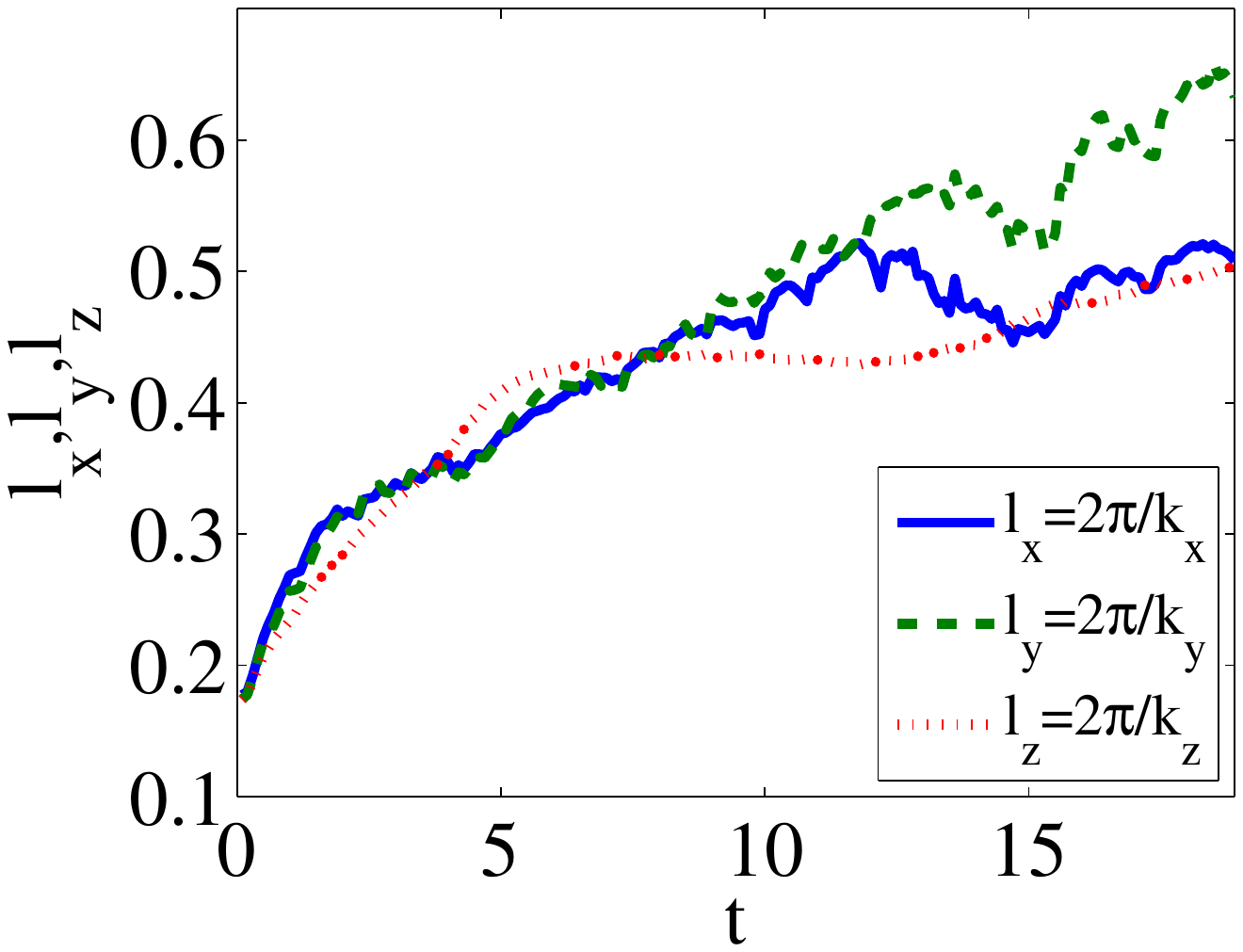}}
		\caption{Typical lengthscales $\ell_x=2\pi/k_x$, $\ell_y=2\pi/k_y$ and $\ell_z=2\pi/k_z$ corresponding to the simulations 3DAniso1 and 3DAniso2 respectively. }
	\label{fig:Nshortlong3d}
\end{figure}

In three dimensions, for long correlation times, the picture is similar to the one in two dimensions: rapid alignment of the domains in the direction of greatest flow amplitude (Figure~\ref{fig:Nshortlong3d}(a)), and eventual saturation of the domain size.
For the case of short correlation times (Figure~\ref{fig:Nshortlong3d}(b)), asymptotic switching of the domain alignment occurs by $t\approx 4$: for $t\apprle 4$ we have $\ell_x\approx \ell_y$ and $\ell_z<(\ell_x,\ell_y)$, while for $t\apprge 4$ the trend reverses, with $\ell_x\approx \ell_y$ persisting, but $\ell_z>(\ell_x,\ell_y)$.  This trend is obscured by $t\approx 8$ where analysis breaks down.  However, by this point, finite-size effects  spoil the scaling laws, since $\ell_x,\ell_y,\ell_z$ are all comparable to half the box size.
The initial phase of the evolution up to $t\approx 4$ can be explained again with respect to Equation~\eqref{eq:sig_eff}, which now reads
\begin{equation}
\sigma(\veck_\perp,k_z)=\mydiff\left(\veck_\perp^2+k_z^2\right)-\gammanondim\mydiff\left(\veck_\perp^2+k_z^2\right)^2-\mathcal{D}_\mathrm{eff}k_z^2,\qquad \veck_\perp=(k_x,k_y).
\label{eq:sig_3D}
\end{equation}
From Equation~\eqref{eq:sig_3D} it therefore follows that any mode $\veck=(\veck_\perp,0)$ will be favoured by the linearized dynamics, meaning that dominant growth of the domains in the $x$ and $y$-directions is equally likely.  Thus, the dynamical model constructed in the two-dimensional case for the short correlation times persists in the 3D case also, albeit that an extra degree of freedom pertains, such that domain alignment in the $x$- and $y$- directions is equally likely in 3D.

\section{Conclusions} 
\label{sec:conclusions}

Summarizing, we have performed a unified and detailed analysis of the phase separation in the presence of chaotic shear flow, for different values of the diffusivity and of the correlation time, in two and three dimensions. It is shown that the correlation time of the random-phase sine flow (itself a simple model of turbulence) can be used to tune the outcome of phase separation: for long correlation times, large flow amplitudes and sufficiently small diffusivities not only is the phase separation arrested and a filamentary concentration field produced, but the concentration variance decays exponentially, reminiscent of advection-diffusion. Flow time scales therefore play an important role, and not just flow amplitudes and lengthscales. In earlier studies (\textit{e.g.} Ref.~\cite{{Berti_goodstuff}}), results were presented as a function of these latter parameters, but the simulations were carried out with a dynamical model of flow (Navier-Stokes equations). As a consequence, the correlation times obeyed the dynamics and might also vary from one case to the other. Our findings are robust to changes in the Cahn number and the dimensionality.

The onset of the diffusive regime (itself characterized by a decay of the concentration variance) is interpreted herein in terms of Batchelor lengthscales, a theory shown to be self-consistent. In the hyperdiffusive regime, the importance of dimensionality is illustrated by investigating the concentration PDF in the ultimate state of phase separation. Turning to anisotropic flows, it is found that the correlation time and the diffusivity can again be used as switches, this time to tune the direction in which the binary-fluid domains align. 
It will be useful to investigate the universality of these results with respect to active Cahn--Hilliard fluids forced in a turbulent manner according to various  protocols in two and three dimensions.

\acknowledgments

L\'ON acknowledges the DJEI/DES/SFI/HEA Irish Centre for High-End Computing (ICHEC) for the provision of computational facilities and support.  L\'ON also acknowledges Hendrik Hoffmann's maintenance of the Orr computer cluster in the School of Mathematical Sciences in UCD and support for users of the same.

\appendix

\section{Measure of domain scale based on structure-function calculations}
\label{app:ellsf}

The calculation of the typical wavenumber $k_1$ is described in detail.  This is the  wavenumber characterising the domain scale and is obtained from structure-function calculations as follows.
The wavenumber $k_1$ is defined as
\begin{equation}
k_1=\frac{\int_0^\infty ks(k,t)\,\mathd k}{\int_0^\infty s(k,t)\,\mathd k}.
\label{eq:k1_slow}
\end{equation}
Here, $s(k,t)$ denotes a spherically-averaged power spectrum.  However, its exact form warrants some further discussion.  First, the following correlation function is obtained:
\[
S(\veck,t)=\frac{1}{L^3}\int\limits_{[0,L]^3}\!\mathd^3 x\!\int\limits_{[0,L]^3}\!\mathd x'\,\mathe^{-\imag\veck\cdot\vecx}\left[C(\vecx+\vecx')C(\vecx)-\langle C\rangle^2\right].
\]
The structure function $S(\vecx,t)$ is normalized and its spherical average computed, to produce
\[
s(k,t)=\frac{1}{(2\pi)^3}\frac{\widetilde{S}(k,t)}{\langle C^2\rangle-\langle C\rangle^2}.
\]
Here, the tilde denotes spherical averaging; the spherical average $\widetilde{\phi}(k)$ of any integrable function $\phi(\veck)$ is defined as
\[
\widetilde{\phi}(k)=\frac{1}{4\pi}\int \mathd\Omega\,\phi(\veck),
\]
where $\mathd\Omega$ is the element of solid angle in three dimensions and the integral is taken over all solid angle.   Also, $C_{\veck}$ is the Fourier transform of $C(\vecx,t)$ and $\widetilde{C}_{\veck}$ is the associated spherically-averaged Fourier transform. For a symmetric binary fluid, $\langle C\rangle=0$, and $s(k,t)$ reduces to the spherically-averaged power spectrum:
\[
s(k,t)=\frac{1}{(2\pi)^3}\frac{|\widetilde{C}_{\veck}|^2}{\langle C^2\rangle}.
\]
In this instance, the computation of $k_1$ dramatically simplifies:
\begin{equation}
k_1=\frac{\int \mathd^3 k\frac{1}{k}|C_{\veck}|^2}{\int \mathd^3 k\frac{1}{k^2}|C_{\veck}|^2}.
\label{eq:k1_fast}
\end{equation}
For postprocessing of the numerical data, this matters: evaluation of the formula~\eqref{eq:k1_fast} involves only summation of array elements and fast-Fourier transforms, while evaluation of the equivalent formula~\eqref{eq:k1_slow} involves spherical averaging, which when done naively, has operation count $O(N^6)$.  Thus, Eq.~\eqref{eq:k1_fast} is used for postprocessing.

\bibliographystyle{unsrt}

\end{document}